\documentclass[usenatbib]{mnras}

\usepackage[T1]{fontenc}
\usepackage{ae,aecompl}

\usepackage{graphicx}	
\usepackage{amsmath}	
\usepackage{amssymb}	
\usepackage{makecell} 
\defcitealias{KT18}{KT18}
\usepackage{bm}
\newcommand{\gradZ}{ \left\langle \nabla Z \right\rangle }

\title[Geostatistics of Galaxies I]{A geostatistical analysis of multiscale metallicity variations in galaxies [I]: Introduction and comparison of high-resolution metallicity maps to an analytic metal transport model}

\author[Metha, Trenti]{
Benjamin Metha$^{1,2}$\thanks{\hbox{methab@student.unimelb.edu.au}}, Michele Trenti$^{1,2}$, Tingjin Chu$^{3}$
\\
$^1$School of Physics, The University of Melbourne, VIC 3010, Australia\\
$^2$Australian Research Council Centre of Excellence for All-Sky Astrophysics in 3-Dimensions, Australia \\
$^3$School of Mathematics and Statistics, The University of Melbourne, VIC 3010, Australia\\
}

\date{Accepted XXX. Received YYY; in original form ZZZ}

\pubyear{2021}

\begin{document}
\label{firstpage}
\pagerange{\pageref{firstpage}--\pageref{lastpage}}
    \maketitle

\begin{abstract}
Thanks to recent advances in integral field spectroscopy (IFS), modern surveys of nearby galaxies are capable of resolving metallicity maps of H{\sc ii} regions down to scales of $\sim 50$pc. However, statistical analysis of these metallicity maps has seldom gone beyond fitting basic linear regressions and comparing parameters to global galaxy properties. In this paper (the first of a series), we introduce techniques from spatial statistics that are well suited for detailed analysis of both small- and large-scale metallicity variations within the interstellar media (ISMs) of local galaxies. As a first application, we compare the observed structure of small-scale metallicity fluctuations within 7 local galaxies observed by the PHANGS collaboration to predictions from a stochastic, physically motivated, analytical model developed by Krumholz \& Ting. We show that while the theoretical model
underestimates the amount of correlated scatter in the galactic metallicity distributions by $3-4$ orders of magnitude, it
provides good estimates of the physical scale of metallicity correlations. We conclude that the ISM of local spiral galaxies is far from homogeneous, with regions of size $\sim 1$ kpc showing significant departures from the mean metallicity at each galactocentric radius.
\end{abstract}



\section{Introduction} \label{sec:intro}

Chemical elements heavier than helium (``metals" hereafter), such as oxygen and iron, are produced by stars during their lives and deaths, and their spatial distribution is therefore inherently linked to both star formation processes and inflow/outflow gas diffusion processes in galaxies (e.g. \citealt{EdmundsGreenhow95, Freeman+Bland-Haworth02, Finlator17, Ma+17, Bresolin19, Sharda+21a}). In particular, the spatial distribution of metals within a galaxy can provide important clues into the galaxy's star formation history, and offers insight into the physical processes that redistribute metals throughout the interstellar (ISM) and intergalactic (IGM) medium, such as inflows of pristine gas from the extragalactic environment \citep{Pace+21}, turbulent motions driven by thermal or gravitational instabilities \citep{deAvillez+MacLow02, Scalo+Emlegreen04, KT18}, interactions with other nearby galaxies \citep{Kewley+10, Torrey+12}, and large-scale streaming motions driven by bars and spiral arms \citep{DiMatteo+13, Grand+16, Ho+17}. Understanding the internal structure of the ISM, and the physical processes that generate and regulate this structure, is one of the greatest challenges in modelling galaxy formation and evolution 
\citep{Naab+17}.

Observationally, measuring the internal metallicity structure of a galaxy can be challenging \citep{Kewley+19, MaiolinoMannucci19}. Metallicity measurements using strong emission line diagnostics require high-fidelity spectroscopic data, limiting surveys to target galaxies that are large and local (but see e.g. \citealt{Yuan+11, Nicha+16,  Wang+19b, Wang+19a} for examples of spatial metallicity studies of gravitationally-lensed, high-redshift galaxies). The most common statistic extracted from these observations is the metallicity gradient: the slope of the trendline fit between metallicity and galactocentric radius.
Most local spiral galaxies have negative metallicity gradients \citep[e.g.][]{Searle71, VilaCostas+92, Zaritsky+94, Berg+13, Berg+20, Ho+15, Belfiore+2017, HP+18}, consistent with inside-out star formation \citep{Boissier+Prantzos99}. It is still an open question whether these metallicity gradients are produced by global galaxy properties such as stellar mass \citep{Ho+15, Belfiore+2017}, or by small-scale versions of well-known galaxy scaling relations, such as the local mass-metallicity relation \citep{Rosales-Ortega13, Erroz-Ferrer+19}.

Using modern, high-resolution integral field spectroscopy (IFS) surveys (e.g. \citealt{CALIFA, MANGA, SAMI}), we now have enough data to go beyond the metallicity gradient, and understand metallicity maps in two dimensions using more advanced statistical techniques. Progress has already been made in this area, e.g. by searching for azimuthal metallicity variations in high-resolution IFS data \citep{Ho+17, Ho+19, Kreckel+19}, but the field is still in its infancy, with different teams using different statistical methodologies and data sets, and arriving at different conclusions.
For example, using data from the CALIFA survey, \citet{Sanchez-Menguiano+19} investigated correlations between local variations in metallicity and the star formation rate (SFR), finding that the two quantities were positively correlated on local scales for high-mass galaxies, yet negatively correlated for low-mass galaxies. Conversely, \citet{Erroz-Ferrer+19} find positive correlations between local SFR density and metallicity for all galaxies in the MAD survey, with no dependence on mass.

Attempts have also been made to use metallicity maps of IFS data to constrain the length scale over which galaxies are chemically well-mixed.
By assuming that no azimuthal metallicity variations should naturally exist and computing deviations from the metallicity gradient, \citet{Sanchez+15} obtained an upper bound on the metal mixing scale of $4.6$ kpc for the galaxy NGC 6754. Using a different methodology, by computing the mean standard deviation around larger and larger sub-regions of PHANGS galaxies, \citet{Kreckel+20} found that metallicities of H{\sc ii} regions stopped being correlated after $\sim 600$ pc. By computing the two-point correlation function in metallicity fluctuations after subtracting a radial trend and fitting an analytical model to 100 galaxy maps measured by the CALIFA survey using Markov Chain Monte-Carlo methods, \citet{Li+21} determined the metallicity correlation length to be $\sim 1$ kpc. These contrasting results come from vastly different methodologies, each with their own set of implicit assumptions about the metallicity structure of galaxies. To contribute to this evolving area of research, we aim to introduce a cross-disciplinary approach to the analysis of metallicity maps using tools that have been developed and extensively validated in the context of geostatistical analysis.

Geostatistics is a sub-field of spatial statistics that focuses on the analysis of spatial and spatiotemporal data over a continuous domain.
Since its inception in the late 1950s \citep{Matheron1954}, it has seen applications over a wide range of disciplines, including epidemiology, meteorology, ecology, soil science, and economics \citep{Wikle+19}.
Geostatistical techniques often go beyond methods of classical statistics in that nearby data points are expected to be correlated, and an emphasis is placed upon separating random fluctuations in spatially varying stochastic data from uncorrelated measurement error. Geostatistics can be used to analyse multi-scale fluctuations in random fields in both real and simulated data, as well as to infer the expected structure of spatially-varying incompletely-sampled data starting from theoretical models, making it a versatile tool for comparing observational data to predictions. However, despite their potential utility, limited applications of these techniques have been seen in the field of extragalactic astrophysics so far. 

In this paper (the first in a series), we aim to introduce key geostatistical techniques in observational astronomy. To demonstrate the usefulness and potential of this approach, we analyse metallicity fluctuations within a sample of galaxies from the PHANGS survey \citep{Kreckel+19}, and compare the results to those expected from the analytical model for metal diffusion developed by 
\citet{KT18} (hereafter \citetalias{KT18}).

The structure of this paper is as follows. In Section \ref{sec:stat_methods}, we introduce a versatile geostatistical framework that can be used to construct a stochastic, spatially-varying model of the ISM, and introduce the \emph{semivariogram}: a mathematical tool that can be used to analyse the size and spatial scale of metallicity fluctuations. In Section \ref{sec:data}, we present the sample of $7$ galaxies to be analysed, and give an overview of how the data products in this work are computed. In Section \ref{sec:NGC2385} we outline our methods for computing semivariograms for the galaxies in our sample before applying it to compare the observed small-scale metallicity fluctuations to the predictions of the \citetalias{KT18} analytical model in Section \ref{sec:model}. We discuss limitations of and extensions to this analysis in Section \ref{sec:discussion}, and present a summary of our main results in Section \ref{sec:conclusions}.

\section{Geostatistical Methods} \label{sec:stat_methods}

As a field of mathematics, geostatistics is young. Born in the 1950s-1960s \citep{matheron1963, cressie93}, it seeks to describe how stochastic processes vary over a continuous spatial or spatio-temporal domain. In its modern form, geostatistics is comprised of a multitude of tools and techniques that can be used to analyse the structure of correlations in measurements occurring over a spatial domain; understand the physical processes responsible for a stochastic spatial phenomenon; and predict/forecast values of natural variables at unmeasured locations. It is clear that there are natural applications of these tools to astronomical datasets. 

While the scope of the subject is vast, in this introductory paper we limit our discussion to only cover the tools that we use in this analysis.
In Section \ref{ssec:heirarchy}, we outline a general geostatistical approach for modelling metallicity fluctuations within the ISM, with an emphasis on separating measurement error from true random deviations from a spatially-varying mean metallicity. In Section \ref{ssec:semivariograms}, we introduce the semivariogram, a mathematical tool which can be used to reveal information about the correlation structure of metallicity variations in both real data and theoretical models.

\subsection{Hierarchical models of spatial structure}
\label{ssec:heirarchy}
Metal production and diffusion in galaxies is an intrinsically stochastic process, which varies over space. We can describe the position-dependent metallicity field 
$Z(\vec{x})$ as a \emph{random field} (for a formal definition, see e.g. \citet{chiles_delfiner99}). 
When we attempt to measure the value of a random field at a point $\vec x$, observations are confounded by measurement error:

\begin{equation}
Z_{\rm obs}(\vec x) = Z(\vec x) + \epsilon(\vec x)
\end{equation}

Here, $Z(\vec x)$ is the true value of the metallicity at the location $\vec x$, $Z_{\rm obs}(\vec x)$ is the measured value of the metallicity, and $\epsilon$ is a random field associated with the measurement error of each observation. Note that $\epsilon$ and $Z$ are both random fields -- that is, there is randomness both in the measurement of the data, and in the quantity being measured itself. To make this problem tractable, we assume that $\epsilon(\vec x)$ and $Z(\vec x)$ are independent of each other at all locations, and that the expectation of the error $E \left[ \epsilon(\vec x) \right] = 0$ for all $\vec x \in D$.\footnote{It is well known that different metallicity diagnostics show large systematic differences from each other -- see, e.g. \citet{Kewley+Ellison08}. These offsets are corrected for when the mean metallicity of the galaxy $\mu$ is subtracted (see Section \ref{sec:NGC2385}), leaving only small, nonlinear uncertainties with mean zero associated with the metallicity diagnostic.
We investigate the possible impact of using different metallicity indicators in Appendix \ref{ap:O3N2}.}

Following \citet{Wikle+19}, we may further break down $Z$ into two components - one nonrandom \emph{process mean} $\mu(x)$, and a spatially varying random component $\eta(x)$ with mean zero:

\begin{equation}
\label{eq:breakdown_true_metallicity}
Z(\vec x) = \mu(\vec x) + \eta(\vec x)
\end{equation}

Under this formulation, spatial statistics then becomes a task of modelling three things: the underlying mean process $\mu$, which may change spatially in a predictable way; the structure of real correlated variance between data points $\eta$; and the structure of (possibly correlated) measurement errors, $\epsilon$.

Historically, the process mean for a galaxy's metallicity field has been assumed to vary linearly with the distance from the galactic centre:

\begin{equation}
    \mu(\vec{x}) = Z_c +  \gradZ \cdot r(\vec{x})
    \label{eq:z_grad}
\end{equation}

where $Z_c$ is the central metallicity, $\gradZ$ is the average metallicity gradient computed using least-squares fitting, and $r(\vec{x})$ is the deprojected distance from the galaxy's centre to each location $\vec{x}$. More recently, the process mean for metallicity has also been considered to vary on other spatially-varying conditions, such as the star formation rate or efficiency of gas (e.g. \citealt{Erroz-Ferrer+19, Sanchez-Menguiano+19, Wang+Lilly21}), or the presence of spiral arms (e.g. \citealt{Ho+17, Sanchez-Menguiano+20}). In this study we will consider the process mean to depend solely on the metallicity gradient, as described in Equation \ref{eq:z_grad}.

Limited research has been done into observational determination of $\eta(\vec{x})$, as the uncertainties associated with metallicity measurement $\epsilon(\vec{x})$ are very high. In the next section, we introduce a graphical technique that can be used to disentangle uncorrelated measurement errors from true random fluctuations in the metallicity field.

\subsection{The Semivariogram}
\label{ssec:semivariograms}

\begin{figure*}
    \centering
    \includegraphics[width=0.98\textwidth]{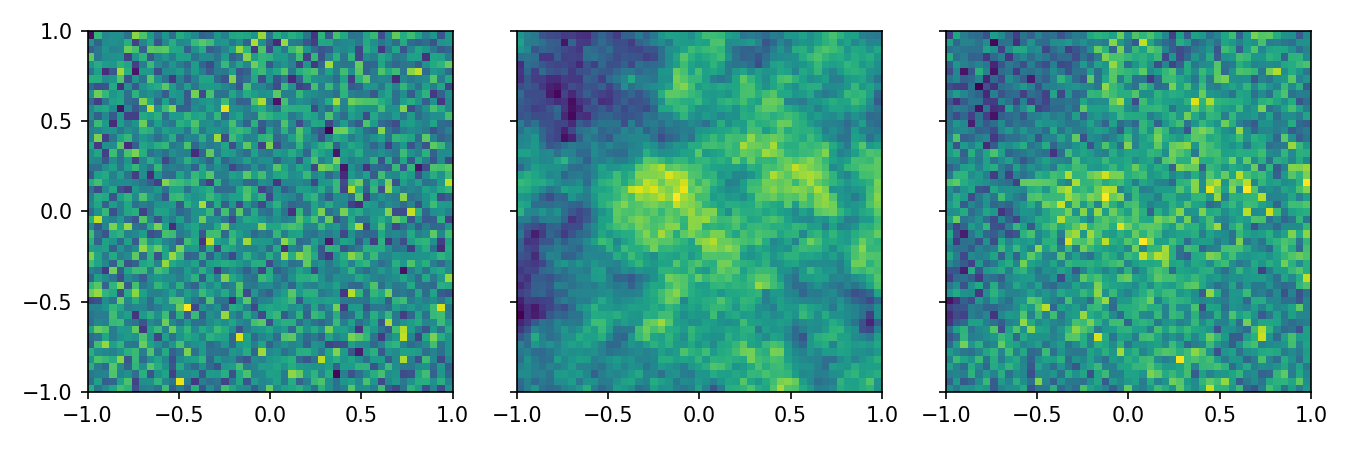}
    \includegraphics[width=0.96\textwidth]{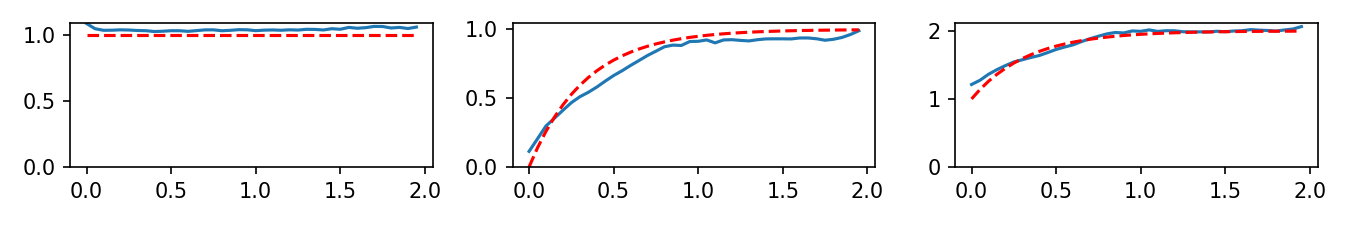}
    \caption{Several realisations of random fields computed from simple covariance functions (top panels), and the semivariograms associated with them (lower panels). \emph{Left:} spatially uncorrelated noise, leading to a flat semivariogram. \emph{Centre:} data is correlated over a short scale ($1/3$ arbitrary units). \emph{Right:} a combination of white noise and correlated spatial variation. The top plots show a realisation of these random correlation functions. In the lower panels, red dashed lines show the theoretical semivariogram function used to generate this data, and blue solid lines show the empirical semivariograms computed from these maps.}
    \label{fig:svg}
\end{figure*}

A key tenet of spatial statistics is Tobler's First Law of Geography, which states that nearby events are more related to each other than those that are further apart \citep{Tobler1970}. The \emph{semivariogram} is a mathematical tool that can be used to assess the degree of similarity between nearby data points. It is defined as follows:

\begin{equation}
    \label{eq:def_svg}
    \gamma(h) = \frac12 \text{Var} \left( Z(\vec{x}) - Z(\vec{y}) \right).
\end{equation}

Here, the variance is computed over all pairs of points $\vec x$ and $\vec y$ for which $h - \delta \leq | \vec x - \vec y| \leq h + \delta$, where $\delta$ is a small, fixed, arbitrary constant.\footnote{The PHANGS datasets analysed in this work have a resolution of $40-70$pc.
Throughout this paper, we choose $\delta h$ to be $100$ pc, allowing us to compute $\gamma(h)$ over bins of width $0.2$ kpc.}

Intuitively, the semivariogram shows how variance between pairs of data points increases as their separation increases. In Figure \ref{fig:svg}, we show the semivariograms associated with several toy models of metallicity fluctuations. In the left panel, all of the variance in the data is uncorrelated, leading to a flat semivariogram. In the middle panel, nearby data points are positively correlated, with correlation coefficient decreasing exponentially with distance, leading to an exponential covariance function: $\gamma(h) = A - A\exp(-h/\alpha)$. Here, $A$ represents the strength of correlated metallicity fluctuations (chosen to be $1$ arbitrary unit), and $\alpha$ is the scale over which metallicities are correlated (chosen to be $1/3$ arbitrary units). In the right panel, we add both distributions together. Visually, it is harder to see the correlated variations due to the white noise: however, with the semivariogram, these two effects are clearly distinguishable. The presence of the uncorrelated white noise simply acts to raise the overall value of the semivariogram without changing its shape. This example illustrates how semivariograms can be used to disentangle uncorrelated metallicity measurement errors (i.e. $\epsilon$) from true small-scale metallicity variations with a predictable correlation structure (i.e. $\eta$), even when the noise is very large.

The semivariogram also satisfies several nice mathematical properties. In situations where the covariance between data points depends only on the distance between them (that is, there exists a positive definite function $C(r)$ such that ${\rm Cov}\left(Z(\vec{x}), Z(\vec{y})\right) = C(r)$ for all points $\vec{x}, \vec{y}$ separated by a distance of $r$), then the following relationship between covariance and the semivariogram holds:

\begin{equation}
    \label{eq:cov_to_svg}
    \gamma(h) = C(0) - C(h)
\end{equation}

This can in turn be used to relate the semivariogram of a random field to its two-point correlation function, assuming that the random field under investigation is stationary and isotropic.

\begin{equation}
    \gamma(h) = \sigma^2\left(1-\xi(h)\right)
\end{equation}

Here, $\xi(h)$ is the Pearson correlation coefficient for the value of $Z$ at two points separated by a distance of $h$, and $\sigma^2$ is the variance of $Z$ throughout the entire random field.

\section{Data} \label{sec:data}

We use data from the Physics at High Angular resolution in Nearby GalaxieS (PHANGS) collaboration.\footnote{\url{www.phangs.org}} Specifically, we consider the metallicity maps of H\textsc{ii} regions for the 7 local galaxies presented in \citet{Kreckel+19} for which ALMA-CO data on the molecular gas density and velocity is available \citep{Sun+20}.\footnote{While metallicity data is also presented in \citet{Kreckel+19} for IC5332, no significant detection of CO(2–1) emission has been observed for this galaxy \citep{Pessa+21}.}.
For convenience, we summarise the data pipeline of the PHANGS-MUSE observation campaign here; for further details, we refer the reader to \citet{Kreckel+19}.

Using the MUSE spectrograph at the VLT, IFU observations are taken of the central star-forming regions of several nearby spiral galaxies, with an angular scale of $0.2\arcsec $/pixel and typical seeing of $0.5\arcsec-1.0\arcsec$, equivalent to a median physical resolution scale of $~50$pc \citep{Kreckel+19}. To make up for the small FOV of the MUSE instrument, several (5-12) pointings are taken of each galaxy, to make a mosaic with large spatial coverage ($\sim 100$ kpc$^2$). 
Using the IDL software package \textsc{LZIFU} \citep{Ho+16}, the stellar continuum is fit to and subtracted from each spaxel, and the strength of the H$\alpha$ emission line is fit.
From these H$\alpha$ intensity maps, H\textsc{ii} regions are isolated from regions dominated by DIG emission using H\textsc{ii}phot \citep{Thilker2000}. 
For each integrated H\textsc{ii} region, the stellar continuum and a collection of strong gas emission lines are fit using GANDALF (Gas AND Absorption Line Fitting; \citealt{Sarzi+06}). 
Dust extinctions are then corrected for using the Balmer decrement, and the extinction law of \citet{Fitzpatrick99}, assuming $R_V=3.1$. 
Regions that are dominated by AGN or shock-driven flux are cut from the data set, using the BPT diagnostics of \citet{Kewley+01} and \citet{Kauffmann+03}. 
Several data quality cuts are applied, including discarding all H\textsc{ii} regions for which S/N is below 5 for any emission line used to compute the metallicity. Together, these cuts remove $\sim 10\%$ of H\textsc{ii} regions.
Finally, for all remaining spaxels, ratios in the fluxes of dust-corrected emission lines are converted into oxygen abundances using the Scal strong-ling calibration, following \citet{Pilyugin+Grebel16}, producing high-resolution metallicity maps for these galaxies.

Global properties of these target galaxies are presented in Table \ref{tab:obs_table}. The mass and SFR for all galaxies are taken from Table 1 of \citet{Kreckel+19}, and distance estimates are taken from \citet{Anand+21}.
Using data from the PHANGS-CO survey \citep{Sun+20}, mean values of the molecular Hydrogen $1D$ velocity dispersion $(\sigma_{H_2,1D})$ and the gas surface density $(\Sigma_{H_2})$ are computed from each galaxy. Similarly, for each H\textsc{ii} region, the electron density $n_e$ is estimated using the [S{\sc ii}] $\lambda$6717/[S{\sc ii}] $\lambda$6731 line ratio and the diagnostic of \citet{Kewley+19b}. We report the mean value of $n_e$ over all H\textsc{ii} regions for each galaxy in Table \ref{tab:obs_table}. As these density values typically fall in the low-density regime, their uncertainties are very high for any individual H\textsc{ii} region. 
Using a bootstrapping method, the uncertainty of the mean value of $n_e$ for each galaxy was calculated by drawing values for the [S{\sc ii}] $\lambda$6717/[S{\sc ii}] $\lambda$6731 line ratio and the metallicity of each H\textsc{ii} region $1000$ times. The median and 95\% confidence interval of each galaxy averaged value of $n_e$ is reported in Table \ref{tab:obs_table}.



\begin{table*}
    \centering
\begin{tabular}{lrrrrrrrrr}
\hline
Name &  Type &    D (Mpc) & PA & $i$ &  $\log_{10}(M_*/M_\odot)$ &  \makecell{SFR  \\ ($M_\odot$ yr$^{-1}$)} &  \makecell{ $\Sigma_{H_2}$ \\ ($M_\odot$ pc$^{-2}$)}    &   \makecell{ $\sigma_{H_2,1D}$ \\(km s$^{-1}$)}  &   \makecell{ $n_e$ \\(cm$^{-3}$) }\\
\hline
NGC 628  & Sc  &   9.84 & 20.9  & 8.7  & 10.2 &   1.82 &  23.56 &  4.20 &  $45.56^{+17.06}_{-4.66}$ \\
NGC 1087 & SBc &  15.85 & 177.3 & 41.3 &  9.8 &   1.12 &  36.46 &  5.89 &  $27.21^{+3.97}_{-3.14}$ \\
NGC 1672 & SBb &  19.40 & 135.7 & 37.5 & 10.2 &   3.02 &  71.65 &  7.32 & $40.45^{+3.30}_{-3.07}$\\
NGC 2835 & SBc &  12.22 & 1.6   & 47.8 &  9.6 &   0.83 &  24.83 &  4.07 &  $33.57^{+3.64}_{-3.15}$ \\
NGC 3627 & SBb &  11.32 & 174.1 & 55.0 & 10.6 &   3.55 &  85.23 &  7.18 & $33.52 \pm 2.37$ \\
NGC 4254 & Sc  &  13.10 & 67.7  & 37.8 & 10.5 &   5.50 &  65.96 &  5.66 & $45.20^{+15.24}_{-3.29}$ \\
NGC 4535 & SBc &  15.77 & 179.8 & 40.7 & 10.4 &   2.24 &  34.85 &  5.14 &  $71.81^{+25.88}_{-8.82}$ \\
\hline
\end{tabular}
    \caption{Global properties of the 7 galaxies investigated in this paper. Distance estimates are taken from \citet{Anand+21}. Hubble type, position angle (PA), inclination ($i$), stellar mass ($M_*$) and SFR values are taken from \citet{Kreckel+19}. Values of $\Sigma_{H_2}$ and $\sigma_{H_2,1D}$ are average quantities computed from the results of \citet{Sun+20}.  Values of $n_e$ are computed for each H\textsc{ii} region using the S\textsc{ii} diagnostic of \citet{Kewley+19b}, and then averaged over all H\textsc{ii} regions.
    }
    \label{tab:obs_table}
\end{table*}

\section{Semivariogram construction} 
\label{sec:NGC2385}

For each galaxy, we compute the deprojected distances between each pair of H{\sc ii} regions using the distance measurements of \citet{Anand+21}. Using the estimated errors in measured metallicities obtained from linear error propagation of line flux uncertainties, a metallicity gradient is fit for each galaxy, using a GLS algorithm from the Python package \texttt{statsmodel}, in order to model $\mu(\vec{x})$. 
We note that this method implicitly assumes that the only deviations that each H\textsc{ii} region has from the mean metallicity in each radial bin originate from measurement errors. Values for the metallicity gradient computed for each galaxy are presented in Table \ref{tab:Z_grads}. These gradients are similar to, yet slightly shallower than, the metallicity gradients calculated in \citet{Kreckel+19}, with a difference of $0.005-0.030$ dex kpc$^{-1}$. This difference may be explained by our decision to use all data points, which may flatten the gradients; whereas in \citet{Kreckel+19}, spaxels within $0.1R_{25}$ are ignored.
 
\begin{table}
    \centering
    \begin{tabular}{l|c|c}
      \hline
       Name& $Z_c$ & $\gradZ$ (dex kpc$^{-1}$) \\
       \hline
       NGC 628  &  $ 8.4858 \pm 0.004$ & $ -0.0139 \pm 0.001$\\
       NGC 1087 &  $  8.3921\pm 0.003$ & $ -0.0218\pm 0.001$\\
       NGC 1672 &  $ 8.4932\pm 0.003$ & $-0.0044\pm 0.000$\\
       NGC 2835 &  $ 8.4804\pm 0.004$ & $-0.0509\pm 0.001$\\
       NGC 3627 &  $ 8.4481\pm 0.004$ & $+0.0030\pm 0.001 $\\
       NGC 4254 & $8.5129\pm 0.002$ & $-0.0082\pm 0.000$\\
       NGC 4535 & $8.5123\pm 0.005$ & $ -0.0063\pm 0.001$\\
      \hline
    \end{tabular}
    \caption{Central metallicity and metallicity gradients for PHANGS galaxies computed via generalised least-squares fitting, with fitting uncertainty reported to 3 decimal palces.}
    \label{tab:Z_grads}
\end{table}

Strong-line metallicity diagnostics are subject to large calibration and/or systematic uncertainties, with different diagnostics producing metallicities that differ by up to $\sim 0.7$ dex \citep{Kewley+Ellison08}. Thus, we subtract off the overall metallicity gradient and only focus on variations around this trend, ensuring that uncertainties associated with the normalisation of these diagnostics are removed. \citet{Kreckel+19} show that the small-scale fluctuations around metallicity gradients obtained from different metallicity diagnostics appear uncorrelated (see their Figure 22 in Appendix C), indicating that differences in normalisation alone cannot account for all of the discrepancy observed between results from different metallicity diagnostics.
Therefore, for robustness, our analysis is repeated using the O3N2 diagnostic together with the calibration of \citet{Marino+13} for each galaxy. A comparison of the semivariograms produced using each metallicity diagnostic is shown in Appendix \ref{ap:O3N2}.

For each galaxy, an empirical semivariogram is computed for the residuals $Z_{\rm obs}(\vec{x}) - \mu(\vec{x})$, using bins of width 0.2 kpc, in order to separate correlated ($\eta(\vec{x})$) and uncorrelated ($\epsilon(\vec{x})$) sources of error (see Figure \ref{fig:svg}). As H\textsc{ii} regions in this data set are defined to be collections of many spaxels, the distance between any pair of H\textsc{ii} regions is larger than both the seeing and the point spread function of MUSE \citep{Kreckel+19}. This justifies our assumption that $\epsilon(\vec{x})$ is not spatially correlated, as observation error in emission line ratios should not be correlated between different H\textsc{ii} regions. 

\subsection{A test case: NGC 2385}

\begin{figure*}
    \centering
    \includegraphics[width=0.9\textwidth]{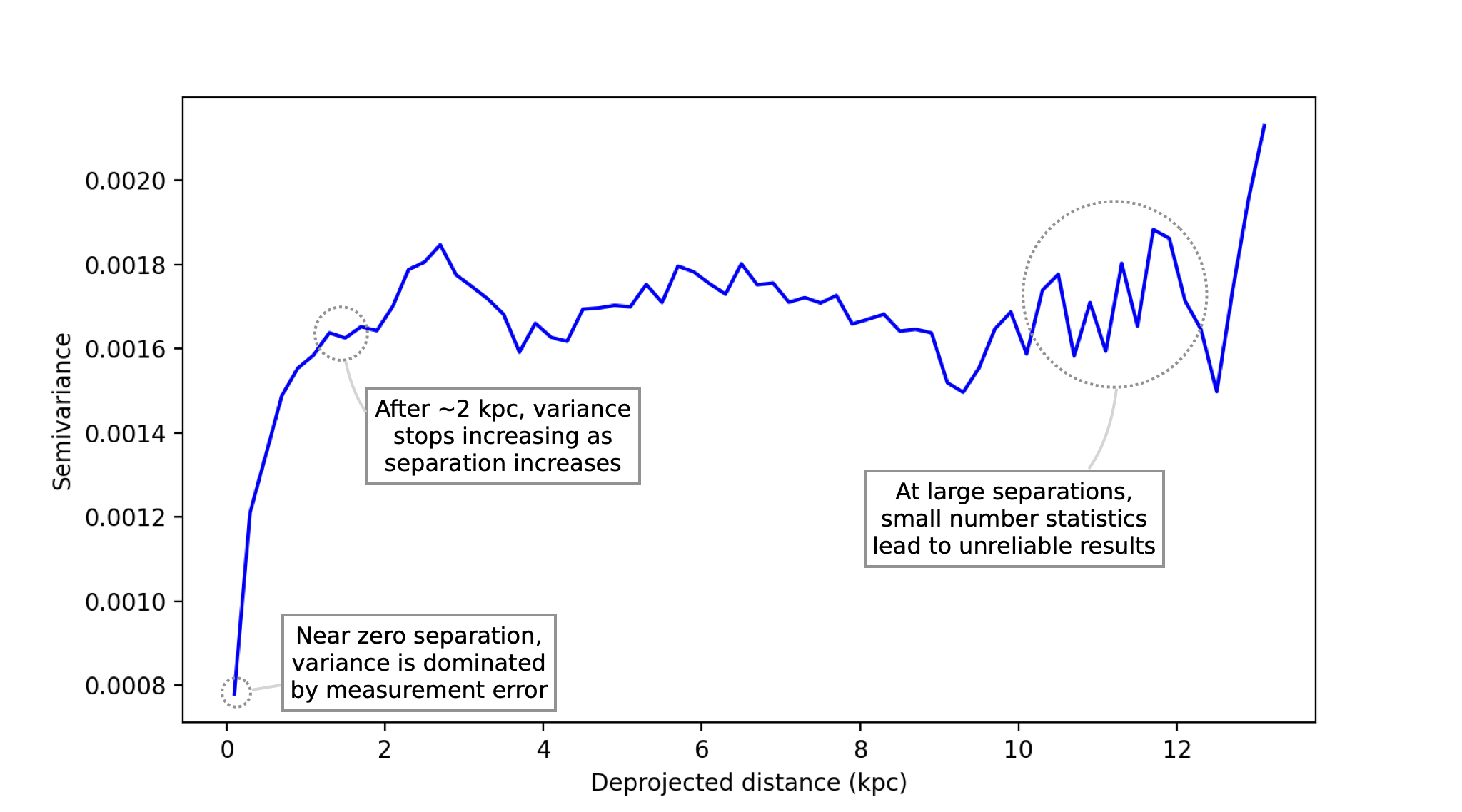}
    \caption{The empirical semivariogram (described by Equation \ref{eq:def_svg}) for the galaxy NGC 2835, with interesting features highlighted. At the smallest separation, the semivariance is dominated by measurement error. At spatial separations below $\sim 1$kpc, metallicities between neighbouring H\textsc{ii} regions are highly correlated.  H\textsc{ii} regions separated by $\gtrsim 2$kpc do not appear to have correlated metallicities, as the semivariogram does not continue to increase beyond this separation. Beyond $\sim 9$kpc, large fluctuations in the semivariance may be attributed to small number statistics, as there are far fewer pairs of H\textsc{ii} regions separated by $\sim 10$kpc than there are separated by $\sim 1$ kpc.}
    \label{fig:one_semivariogram}
\end{figure*}

Figure \ref{fig:one_semivariogram} shows the empirical semivariogram for one PHANGS galaxy, NGC 2835, highlighting the key features that can be observed. The semivariance is smallest between H\textsc{ii} regions that are separated by $0.1 \pm 0.1$kpc, and increases monotonically until a separation of $1.9 \pm 0.1$kpc is reached. Beyond this point, the semivariance is almost constant, showing variations on the $10\%$ level. This indicates that in this galaxy, metallicity variations fail to be positively correlated for H\textsc{ii} regions that are separated by a distance larger than $\sim 2$ kpc, setting an upper limit for the mixing scale of the ISM. 

In the smallest separation bin, a semivariance of $7.79\times10^{-4}$ is seen. This is comparable to the amount of variance expected to be produced by measurement errors of emission-line flux ratios alone: for this galaxy, the mean uncertainty in metallicity measurements reported in \citet{Kreckel+19} is $0.029$ dex, which would add a semivariance of $4.35\times10^{-4}$ to all spatial bins. From this, we can infer that the greatest source of fluctuations in the metallicity map for this galaxy on spatial scales below $200$pc is measurement error. This also provides an upper-limit to the size of non-linear errors in the Scal metallicity diagnostic, as the height of the semivariogram in the smallest bin cannot be smaller than the contribution from measurement error.

This example highlights one interesting feature of the semivariogram method. Because the semivariogram allows one to disentangle observational error from true random fluctuations, it can be used not only to understand small-scale fluctuations $\eta(\vec{x})$ without the confounding effects of noise $\epsilon(\vec{x})$, but also to study the overall size of the measurement uncertainty $\epsilon(\vec{x})$ without needing to model the small-scale variance $\eta(\vec{x})$. One application of this technique is shown in Appendix \ref{ap:O3N2}, in which we recover the result that the uncertainties associated with the O3N2 metallicity diagnostic \citep{Marino+13} are larger than those associated with the Scal diagnositc \citep{Pilyugin+Grebel16}.


\subsection{Full sample}

\label{ssec:full_sample}
The semivariograms for the other six galaxies in the sample are shown in Figure~\ref{fig:theory_svgs}, and are qualitatively similar to that of NGC 2835. All semivariograms flatten out between 0.9 and 3.1 kpc, in good agreement with the findings of \citet{Kreckel+20}. No trends are seen between the separation at which the semivariance no longer increases, and global galaxy properties such as mass, gas velocity, or star-formation rate. Using a sample of 100 local galaxies from the CALIFA survey, \citet{Li+21} found a weak positive correlation ($\rho=0.36 \pm 0.07$) between the metallicity mixing scale and the galaxy size, measured by $R_{25}$. The fact that we do not recover this correlation in our data is likely due to the small number of galaxies analysed in this work.

For NGC 1672, NGC 4254, and NGC 4535, the amount of variance seen between H\textsc{ii} regions separated by less than $0.2$kpc is consistent with the amount of variance expected from measurement error, indicating that the ISM is reasonably homogeneous for these galaxies on scales below $0.2$kpc. For all other galaxies in the sample, the semivariogram at the smallest separation was 2-3 times higher than would be expected from measurement uncertainty alone. This may either be caused by the presence of additional metallicity structures within these galaxies on scales below $0.2$kpc, or from non-linear errors in the Scal diagnostic. Future surveys, such as SDSS-V/LVM \citep{SDSS-V} which has been designed to observe the internal structure of H\textsc{ii} regions in the Milky Way and the Local Group, will have sufficient resolution to determine which of these theories is correct.

When metallicity gradients are computed using standard least-squares methods, all of the variance around the mean trend line is implicitly assumed to originate from uncorrelated measurement errors in the metallicity data. By comparing the semivariogram at small separations to the value of the semivariogram after it has levelled off, we can compute the amount of large-scale variance that instead originates from small-scale ($\sim 1$ kpc) metallicity fluctuations. We find that, for all galaxies in this sample, the amount of variance originating from correlated metallicity structures is approximately equal to the amount of variance originating from uncorrelated error sources. Therefore, modelling the correlation structure of small-scale random metallicity fluctuations is an important consideration for robust metallicity gradient determinations.

\section{Analytical model comparison} \label{sec:model}

In the previous section, we have shown that using high-resolution IFU data, the semivariogram can be used to measure the amplitude and correlation structure of $\eta(\vec{x})$. This allows the predictions of theoretical models to be tested against real data. Here, we demonstrate this process using the theoretical model introduced in \citetalias{KT18}, which we describe below.

\subsection{The \citetalias{KT18} model}
\label{ssec:model_description}

In order to compute an analytical prediction for the two-point correlation function of metallicities throughout a galaxy, \citetalias{KT18} model the small-scale spread of metals as \emph{stochastically-forced diffusion}. Diffusion is modelled as being linear, with its strength regulated by a diffusion coefficient $\kappa$. This is a significant oversimplification of turbulent transport; however, it ensures that the differential equations involved have analytical solutions.

Over time, metals are injected into the ISM via clusters of Type II supernovae. These injection events are modelled as being instantaneous in time and Gaussian in space, with width $x_0$. The total amount of metals added to the ISM by each injection event, $m_{X,i}$, is allowed to vary. 

Following \citetalias{KT18}, we define $\Sigma_X$ to be the surface density of a metal of species $X$ in the ISM. Then, under this model, $\Sigma_X(\vec{x},t)$ is fully described by the stochastic partial differential equation:

\begin{equation}
\frac{\partial}{\partial t} \Sigma_X = \kappa \nabla ^2 \Sigma_X + \sum_{i} m_{X,i} \frac{1}{2\pi x_0^2} \exp \left( -  \frac{\left| \vec{x} - \vec{x_i} \right|^2 }{2x_0^2} \right) \delta(t-t_i)
\end{equation}

Here, the number of injection events over a time period $T$, for a patch of the galaxy with area $A$, is drawn from a Poisson distribution with mean equal to $\Gamma A T$, where $\Gamma$ is the event rate density for injection events. In this model, $\Gamma$ is assumed to be constant throughout space and time: hence, all instances of $\vec{x_i}$ and $t_i$ are drawn from uniform distributions over $A$ and $T$, respectively.

Employing several reasonable approximations, \citetalias{KT18} compute the two-point correlation function for $\Sigma_X$ to be:

\begin{equation}
\xi(x)=\frac{2}{\ln \left(1+\frac{2 \kappa t_{*}}{x_{0}^{2}}\right)} \int_{0}^{\infty} \mathrm{e}^{-x_{0}^{2} a^{2}}\left(1-\mathrm{e}^{-2 \kappa t_{*} a^{2}}\right) \frac{J_{0}(a x)}{a} \mathrm{~d} a
\end{equation}

Here, $J_0$ is the zeroth order Bessel function of the first kind, and $t_*$ is the time over which star formation has been occurring (in this model, a constant star formation history is assumed). The correlation coefficient can be related to the covariance between two events using the following equation:

\begin{equation}
{\rm Cov} \left( \Sigma_X(\vec{x}, t) , \Sigma_X(\vec{x} + \vec{x}', t) \right) = \sigma^2_{\Sigma_X} \xi \left( \left| \vec{x}' \right| \right)
\end{equation}

To understand this equation, it is helpful to consider the two limiting cases of $\xi(r)$. As $r\to 0 $, $\xi(r) \to 1$ -- that is, every data point is perfectly correlated with itself, and the expected covariance of $\Sigma_X$ is simply the expected variance in $\Sigma_X$. Similarly, as $r\to \infty$, $\xi(r) \to 0$, and every pair of data points that are sufficiently separated become completely uncorrelated.

Finally, we convert this covariance in metal surface density, $\Sigma_X$, to covariance in the measured metallicity. These two quantites are related, as $Z = \log_{10}(\Sigma_X / \Sigma_g) + 12$. Using Taylor expansions for the moments of functions of a random variable and assuming that $\Sigma_g$ is constant throughout a galaxy, we arrive at the following approximation for the covariance of the metallicity between data points as a function of their separation, $r$:

\begin{equation}
C(r) = \frac{1}{\ln (10) ^2} \ln \left[ 1 + \frac{\sigma^2_{\Sigma_X}}{\mu^2_{\Sigma_X}} \xi(r) \right] 
\label{eq:cov}
\end{equation}

This approximation will be valid whenever $\sigma_{\Sigma_X} << \mu_{\Sigma_X}$:
\citetalias{KT18} asserts that this is always the case for the physical conditions found in galaxies.

Equation~\ref{eq:cov} is the key ingredient from the \citetalias{KT18} we use to compute the intrinsic covariance between any pairs of points in our geostatistical model. To use this equation, we need to know the ratio between the variance of $\Sigma_X$ and its mean squared. Because this ratio is insensitive to linear rescalings, it is simpler to compute this from the mean and variance of the dimensionless quantity $S_X$, which is equal to $\Sigma_X$ up to a rescaling. Equations for the mean and variance of this quantity are given in \citetalias{KT18} (Equations 37 and 38):

\begin{eqnarray}
\mu_{S_X} &=& t_* \sqrt{\kappa\Gamma} \\
\sigma^2_{S_X} &=& \frac{1+\sigma_w^2}{8\pi} \ln \left[ 1 + \frac{2t_*\kappa}{x_0^2} \right]
\end{eqnarray}

$\sigma_w^2$ is the final parameter required for our covariance function. This parameter is the variance in the mass of metals injected from a single event, divided by the square of the expected mass of metals injected from an event.

\subsection{Model parameter estimation for PHANGS galaxies}
\label{ssec:model_params}
In total, the \citetalias{KT18} model contains five independent parameters: $\Gamma, \kappa, \sigma^2_w, x_0,$ and $t_*$. All of these parameters have a physical meaning, and can be estimated from the global galaxy properties presented in Table \ref{tab:obs_table}.\footnote{We note that many of these parameters are not expected to be constant throughout a galaxy. In particular, the star formation rate $\Gamma$ and the star formation timescale $t_*$ are expected to be higher near the centre of the galaxy. One way to account for these radial trends is to fold these effects into the model for $\mu(\vec{x})$, the expression for the mean metallicity as a function of position. However, such considerations are beyond the scope of this analysis.} A full list of these parameters for each galaxy is presented in Table \ref{tab:KT18_params}. We outline the equations used to estimate these parameters below.

\begin{table}
    \centering
\begin{tabular}{lrrccc}
\hline
Name &  \makecell{$t_*$ \\ (Gyr)} & \makecell{$\kappa$ \\ (pc km s$^{-1}$)} &  \makecell{$\Gamma$  \\ (pc$^{-2}$ Myr$^{-1}$)}& $\sigma^2_w$ &    \makecell{$x_0$ \\ (pc)} \\
\hline
NGC 628  &  8.71 & 728.6 & 1.06e-05 &    20 & 17.2 \\
NGC 1087 &  5.62 & 805.3 & 1.61e-05 &    20 & 18.6 \\
NGC 1672 &  5.25 & 3662.9 & 2.74e-05 &    20 & 15.0 \\
NGC 2835 &  4.79 & 3144.0 & 7.92e-06 &    20 & 19.5 \\
NGC 3627 & 11.22 & 3413.7 & 4.90e-05 &    20 & 16.3 \\
NGC 4254 &  5.75 & 1011.7 & 2.81e-05 &    20 & 16.0 \\
NGC 4535 & 11.22 & 3571.6 & 1.10e-05 &    20 & 13.7 \\
\hline
\end{tabular}
    \caption{Parameters of the \citetalias{KT18} model used for each galaxy, computed from the values listed in Table \ref{tab:obs_table}, using the equations outlined in Section \ref{ssec:model_params} }
    \label{tab:KT18_params}
\end{table}

\subsubsection{Injection rate $(\Gamma)$}
\label{ssec:gamma}

When tracing the oxygen enrichment of a galaxy over time, an injection event is defined to be a collection of Type II supernovae originating from the same star cluster, occurring at the same time. In line with this definition,  $\Gamma$ is taken to be the star cluster formation rate density of the galaxy: $\Gamma = \dot \Sigma_* / M_{cl}$. This definition saves modelling the spatial correlation structure of supernovae. By integrating a cluster mass function that follows a power law with an index of $-2$ \citep{Elmegreen06}, \citetalias{KT18} estimate the mean mass of a cluster to be $M_{cl} = 690M_\odot$, and we adopt this value in this work. The mean star formation rate density for each galaxy is computed from the star formation rate density maps presented in \citet{Pessa+21}. 

\subsubsection{Injection width, $x_0$}
\label{ssec:x0}

Regions of gas produced by expanding supernovae will begin to mix turbulently with the surrounding gas when the outflow velocity of the supernova becomes roughly equal to the ISM velocity dispersion \citep{deAvillez+MacLow02}. The radius at which this occurs for a Type II supernova of energy $10^{51}$ ergs, in an ISM with a density of $n_H$ cm$^{-3}$ and a gas velocity of $\sigma_{g,1} \times 10$ km/s is given by \citep{Draine11}:

\begin{equation}
x_{0} \approx 67 n_{H}^{-0.37} \sigma_{\mathrm{g}, 1}^{-2 / 5} \mathrm{pc}
\label{eq:inj_width}
\end{equation}


\subsubsection{Star formation timescale ($t_*$)}
\label{ssec:t_sf}

This parameter represents the timescale over which stars has been forming. Because the \citetalias{KT18} model assumes galaxies have a constant star-formation history, we may approximate this by simply dividing the stellar mass of each galaxy by its present-day star formation rate: $t_* = M_*/SFR$. Implications of this assumption, and its effects on the predicted metallicity covariance function, are discussed in Section \ref{sec:discussion}.

\subsubsection{Normalised injection mass variance $(\sigma^2_w)$}
\label{ssec:sigma_w}

In \citetalias{KT18}, this is computed from the variation in star cluster masses, assuming that the star cluster mass function follows a power law with a slope of $-2$, following e.g. \citet{Bastian+12, Fall+Chandar12, Adamo+17}. Using this value, \citetalias{KT18} estimate $\sigma^2_w \approx 20$. The authors find that using a different slope, such as $-1.7$ \citep{Murray+10}, does not change this quantity very much ($\sigma^2_w \approx 15$). Because this factor does not appear in the integral used to calculate $\rho(r)$, this factor will only affect the normalisation of the metallicity covariance function in a linear way. For this reason, we do not attempt to fit this parameter to the data.

We note that this assumes that all of the variance in the mass of metals comes from variance in the mass of clusters; that is, all stars are expected to have the same yield. Different clusters will likely have different populations of supernova progenitors. Accounting for this second-order effect would slightly increase the value of $\sigma_w^2$; however, properly accounting for this effect is outside of the scope of this study.

\subsubsection{Diffusion coefficient ($\kappa$)}
\label{ssec:kappa}

Turbulence, powered by gravitational and thermal instabilities and stellar feedback, acts to enhance concentration gradients in the ISM that catalyse molecular diffusion \citep{Pan_Thesis}. While turbulence does not smooth our ISM inhomogeneities on its own, its effect on enhancing $\kappa$ can be estimated at the order-of-magnitude level from global galaxy properties. 

According to \citet{Karlsson+13}, $\kappa$ can be estimated from the largest scale of turbulent eddies, multiplied by the turbulent velocity, divided by $3$. Following \citet{deAvillez+MacLow02}, we can equate the turbulent velocity to the RMS velocity of the gas, $\sigma_g$. \citetalias{KT18} suggest that the scale height of the ISM, $h$, sets the outer scale for turbulence, so $\kappa$ may be approximated as:

\begin{equation}
\label{eq:kappa}
\kappa \approx \frac{h \sigma_g}{3}
\end{equation}

Therefore, in order to determine the strength of turbulence, the $3D$ velocity of the gas phase of the ISM and the scale height of the ISM must both be computed from available data for these galaxies. We discuss how this may be done below.

Firstly, we convert the $H_2$ velocity dispersion along the line-of-sight observed by \citet{Sun+20} into a 3D velocity dispersion assuming isotropy, using the formula $\sigma^2_{H_2} = 3\sigma^2_{H_2,1D}$. Next, we must also account for the fast-moving atomic phase of the ISM. Based on the findings of \citet{Marasco+17}, \citet{Bacchini+19} model the velocity dispersion of $HI$ as being directly proportional to the velocity dispersion of $H_2$ with $\sigma_{HI}/\sigma_{H_2} = 2$; here, we adopt this approach.

Finally, we estimate the fraction of molecular gas, $f_{mol} := \frac{\Sigma_{H_2}}{\Sigma_{HI}+ \Sigma_{H_2}}$, for each galaxy. Using data from The H\textsc{i} Nearby Galaxy Survey (THINGS, \citealt{Bigiel+08}), \citet{Leroy+08} found that $\Sigma_{HI}$ is approximately constant for spiral galaxies up to a distance of $\sim 1 R_{25}$, with $\Sigma_{HI} \approx 6 M_\odot$ pc$^{-2}$. We assume this value to derive first-order estimates of $f_{mol}$ for all galaxies studied in this paper, with results reported in Table \ref{tab:ISM_data}. Then, the derived $f_{mol}$ is used to compute the mass-weighted average 3D gas dispersion velocity for each PHANGS galaxy:

\begin{eqnarray}
\sigma_g  &:=& f_{mol}\cdot \sigma_{H_2} + (1-f_{mol})\cdot \sigma_{HI} \\
 &\approx& f_{mol}\cdot \sigma_{H_2} + (1-f_{mol}) \cdot 2\sigma_{H_2} \nonumber \\ 
 &=& (2-f_{mol})\cdot \sigma_{H_2}
 \label{eq:gas_vel_model}
\end{eqnarray}

Estimates of the ISM velocity dispersion computed using this method are reported for each galaxy in Table \ref{tab:ISM_data}, ranging from $8.4-13.7$ km s$^{-1}$ with an average velocity of $10.9$ km s$^{-1}$, in excellent agreement with results from THINGS \citep{Tamburro+08}. 

For face-on spiral galaxies, the gas scale height, $h$, cannot be measured directly. Instead it must be inferred using a model of galaxy structure. By assuming the density profile of a disc galaxy to be exponential in the $z$-direction with a constant scale height\footnote{This is known to be a simplifying assumption. Studies of gas discs have shown that there is significant flaring of scale height with radius; see e.g. \citet{Bacchini+19} for a discussion. However, in the \citetalias{KT18} model, no flaring is accounted for; so attempting to model this effect for these galaxies is neither necessary nor useful for finding reasonable model parameters.} and the distance from spiral arms to the galactic centres to be equal to the forbidden radius for density waves, \citet{Peng88} describe a method by which $h$ can be calculated from the shape of the spiral arm structure of face-on galaxies by solving the 3-dimensional Poisson equation, and apply this method to 4 local face-on spirals, including NGC 628. Later, \citet{Ma+98} extended this analysis to 71 additional northern spiral galaxies, including NGC 2835 and NGC 4535. We adopt their published values of $h$ for these galaxies. 
For the other 4 galaxies in our sample, in order to have $h$ estimated in a consistent way, we use the results of \citet{Ma2000}, who publish the mean flatness ($h/R_{25}$) computed using the method of \citet{Peng88} as a function of Hubble type. For these galaxies, we take the uncertainty of $h$ to be the dispersion in flatness found in the population of galaxies studied by \citet{Ma2000}. Estimates of $h$ found using this method range from $0.2-1$ kpc, in good agreement with typical gas scale heights determined for local spiral galaxies \citep{Patra+19} and the Milky Way \citep{Carroll+Ostlie07}. Scale heights and values of $R_{25}$ are reported in Table \ref{tab:ISM_data} together with their uncertainties for all galaxies.




\begin{table}
    \centering
    \begin{tabular}{l|c|r|r|r}
      \hline
       Name & $f_{mol}$ & $\sigma_g$ (km s$^{-1}$) & $R_{25}$ (kpc) & $h$ (pc) \\
       \hline
       NGC 628  & 0.80 & 8.74  & 14.31 &$250^\dagger \pm 68$ \\
       NGC 1087 & 0.86 & 11.64 &  6.92 &$207\pm 104$ \\
       NGC 1672 & 0.92 & 13.66 & 17.49 &$805 \pm 245$ \\
       NGC 2835 & 0.81 & 8.42  & 11.37 &$1120^\ddagger \pm 82$ \\
       NGC 3627 & 0.93 & 13.26 & 16.79 &$773 \pm 235 $ \\
       NGC 4254 & 0.92 & 10.62 &  9.53 &$286 \pm 143 $ \\
       NGC 4535 & 0.85 & 10.20 & 18.81 &$1050^\ddagger \pm 105$\\
      \hline
    \end{tabular}
    \caption{Additional ISM properties for the PHANGS galaxy sample. $f_{mol}$ is computed using values of $\Sigma_{H_2}$ from Table \ref{tab:obs_table}, assuming $\Sigma_{H1}=6M_\odot$ pc$^{-2}$ for all galaxies. $\sigma_g$ is computed using Equation \ref{eq:gas_vel_model}. Values of $R_{25}$ are taken from \textsc{HYPERLEDA} and converted into physical distances using the distance estimates of \citet{Anand+21}. References for $h$: ($\dagger$) \citet{Peng88}; ($\ddagger$) \citet{Ma+98}. All other values of $h$ are inferred from the average flatness of galaxies of the same Hubble type published in \citet{Ma2000}.}
    \label{tab:ISM_data}
\end{table}

\subsection{Observed vs Theoretical Semivariograms} \label{ssec:v_theory}

\begin{figure*}
    \centering
    \includegraphics[width=0.9\textwidth]{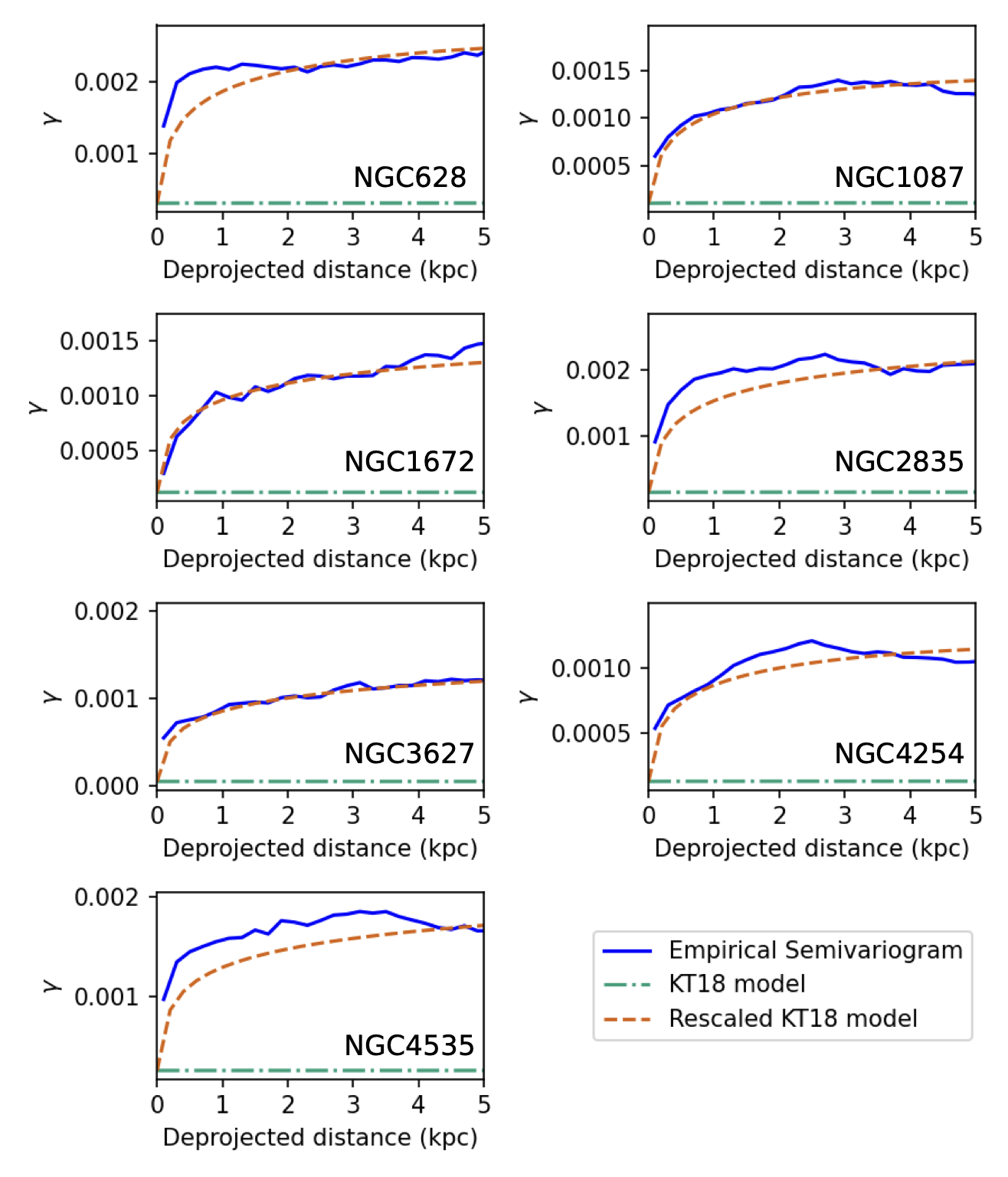}
    \caption{Empirical semivariograms for 7 PHANGS galaxies (blue solid lines) up to a distance of $5$kpc, compared to theoretical semivariograms computed from the model of \citetalias{KT18} using global galaxy properties (dash-dotted green lines). For clarity, rescaled versions of these plots are also shown (dashed orange lines).}
    \label{fig:theory_svgs}
\end{figure*}

Following the methods outlined in Section \ref{sec:NGC2385}, empirical semivariograms are computed for each galaxy from the metallicity map of H{\sc ii} regions, and compared to the theoretical semivariogram obtained from the model of \citetalias{KT18}. We show these results in Figure \ref{fig:theory_svgs}. 

Solid blue lines represent the empirical semivariograms for these data, and are described in Section \ref{ssec:full_sample}.
Green dash-dotted lines show the theoretical semivariograms predicted from the \citetalias{KT18} model, with an added component equal to the expected contribution to the semivariogram from uncorrelated observational error ($\epsilon(\vec{x})$).
From this, we can see that the \citetalias{KT18} model underpredicts the amount of correlated variance in the data by $\sim 3-4$ orders of magnitude.
For clarity, rescaled versions of these plots, where the predicted amount of variance is multiplied by a constant factor, are overplotted as orange dashed lines. The best-fit constants for rescaling are found using the python package \texttt{scipy.optimize}, and
range from a factor of $606$ for NGC 4254, to $1.48\times10^{4}$ for NGC 3627.\footnote{Our result uses a value of the total metallicity scatter as predicted by the \citetalias{KT18} model in an implementation that is different to what was originally published, following private communications with the authors of that paper. This is due to an error in Equation 106 of the original paper, resulting in a published prediction for the variance due to random fluctuations that is too large by a factor of $t_*\sqrt{\Gamma\kappa}$ (Krumholz 2021, private communication). If the originally published form of this equation were to be used, the total variance of metallicity predicted by the model would at face value agree to within a factor of two with the inference from the semivariogram analysis.}

While the total amount of variance predicted by the \citetalias{KT18} model is far less than what is observed, the scale over which metallicities are correlated and the overall shape of the predicted semivariograms show very good agreement to the data, especially for NGC 1087, NGC 1672, and NGC 3627. This agrees with the results of \citet{Kreckel+20}, who found that the correlation scale for small-scale metallicity fluctuations in the PHANGS sample of galaxies was well matched by the predictions of the \citetalias{KT18} model, tuned to the parameters of the Milky Way. A detailed comparison to their work is presented in Section \ref{sec:discussion}.
For other galaxies such as NGC 628, NGC 2835, and NGC 4535, the semivariogram is larger at the small-scale end than is predicted by the \citetalias{KT18} model, indicating that these galaxies contain more inhomogeneities on scales of $\sim 1$ kpc than predicted by the model.

One possible explanation for this effect is that the metallicity is being affected by the presence of spiral arms. Typical arm widths for grand design spirals are $3.3 \pm 1.2$ kpc \citep{Savchenko+20}, and the presence of spiral arms has been found to affect the metallicity distributions of galaxies, both in simulations (e.g. \citealt{DiMatteo+13, Grand+16, Bellardini+21}; but see also \citealt{Molla+19}), and in observations \citep[e.g.][]{Ho+17, Ho+18, Sanchez-Menguiano+20}. To test this, the model for the mean metallicity throughout a galaxy described in Equation \ref{eq:z_grad} could be altered to accommodate this large scale effect by including another covariate, $\mathcal{S}(\vec{x})$ that equals $1$ if $\vec{x}$ is within a spiral arm, and $0$ otherwise:

\begin{equation}
    \mu(\vec{x}) = Z_c +  \gradZ \cdot r(\vec{x}) + \beta_S \cdot \mathcal{S}(\vec{x})
    \label{eq:z_grad_w_spiral_arms}
\end{equation}

Here, $\beta_S$ is a parameter describing the increase of metallicity due to being within a spiral arm. Analogous to $\gradZ$, this could be fit for each galaxy using GLS, or a similar method. 

Environmental masks showing the location of the spiral arms for the sample of galaxies studied in this work are yet to be released by the PHANGS collaboration (Querejeta et. al. 2021 in prep.). Therefore, we postpone this analysis for a future study, pausing here to highlight the fact that while the model we use in this study is basic, the methodology presented in this paper is flexible enough to account for any number of additional second-order effects.


\begin{figure*}
    \centering
    \includegraphics[width=0.98\textwidth]{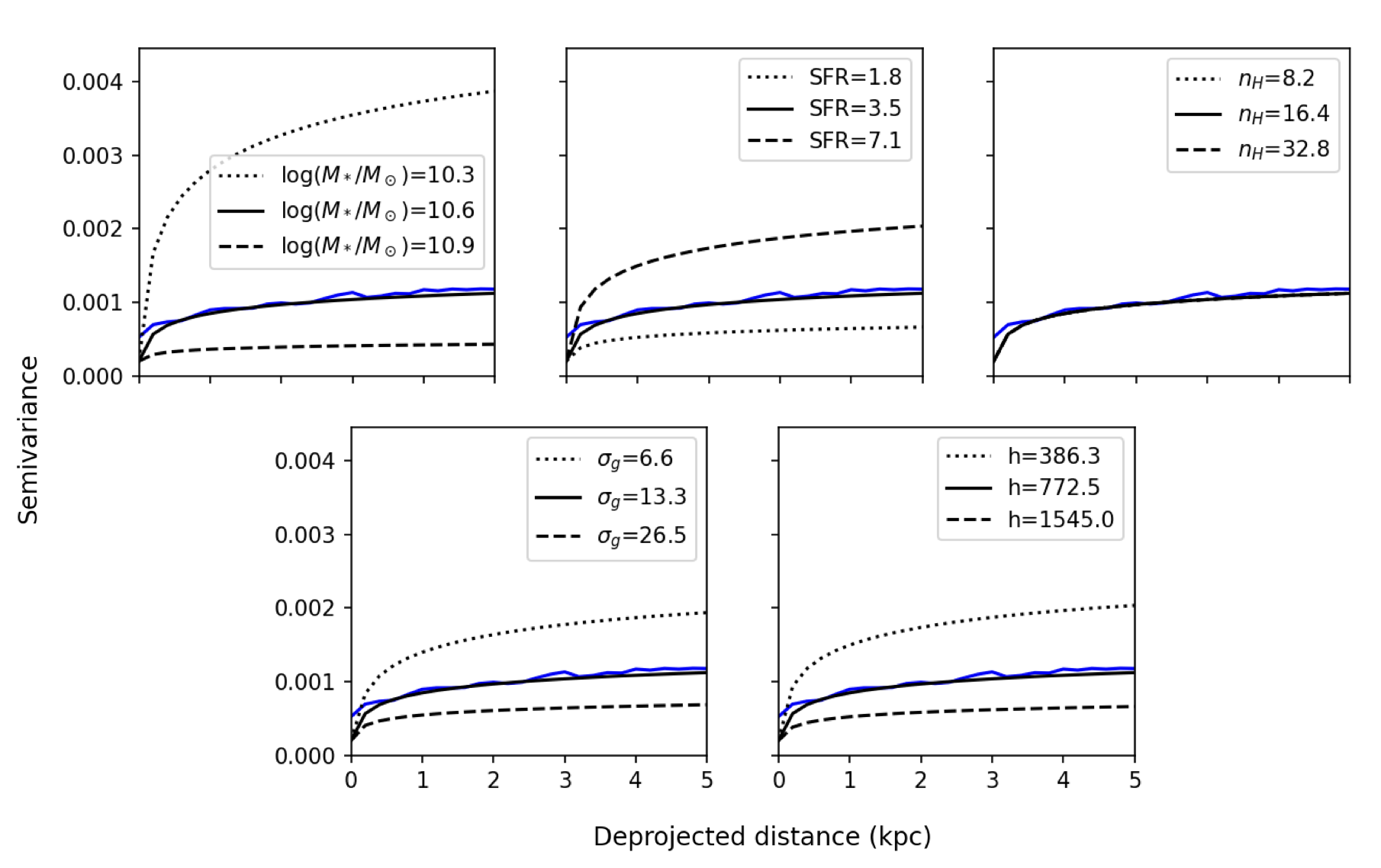}
    \caption{Dependence of the theoretical semivariogram on the measured global properties of NGC 3627, as
    stellar mass (\emph{top left}), star formation rate (\emph{top middle}), hydrogen density (\emph{top right}), gas velocity (\emph{bottom left}), and scale height (\emph{bottom right})
    are altered by a factor of 2. To show the increase in the normalisation constant as any of these parameters are changed, all theoretical semivariograms are rescaled by the same factor, which is calibrated by fitting the fiducial model (solid black line) to the data.}
    \label{fig:obs_sens}
\end{figure*}

\section{Discussion} \label{sec:discussion}

At first glance, it is not clear whether the disagreement between the predictions of the \citet{KT18} model and the data is caused by measurement error in global galaxy properties, or by modelling approximations.
In Figure \ref{fig:obs_sens}, we show how errors in the measured values of global galaxy properties would affect the theoretical semivariograms computed for NGC 3627; results are similar for all galaxies. For clarity, only the rescaled semivariogram is shown in this plot. We show that increasing any of $M_*$, $h$, or $\sigma_g$ acts to flatten the semivariogram; increasing SFR, acts to raise the size of the semivariogram; and changes to $n_H$ of the order of 1 magnitude have a negligible effect. None of these changes significantly alter the range at which the semivariogram flattens out, 
making the result that the spatial scales of metallicity fluctuations observed in the PHANGS galaxy sample match the predicted spatial scales highly insensitive to measurement errors. 
On the other hand, errors in the global properties of the galaxies would need to be very large to account for the discrepancy between the height of  the predicted (theoretical) and observed (empirical) semivariograms. Any discrepancy would also need to be systematic, in order to increase the variance seen in all galaxies rather than decrease it. We therefore conclude that the difference between the modelled and observed semivariograms is unlikely to be caused by measurement errors in global galaxy properties.

One way to increase the variance in metallicity predicted by the \citetalias{KT18} model is by introducing AGN-powered galactic winds. If galactic winds act to constantly remove metals over time, and are uncorrelated with the sites of metal injection (as is the case for AGN-driven galactic winds), then they will act to decrease the mean metallicity of the galaxy. Such winds will not significantly decrease the variance of metal density within the galaxy, as this is dominated by recent metal enrichment events. Together, these two effects would act to increase the variance in $\log(O/H)$. However, if the galactic winds are instead correlated with the sources of metals, as is the case for Type II supernovae driven winds, then the stellar winds will act to both reduce the mean metal density and its variance, resulting in no change in the variance of $\log(O/H)$. Of the seven galaxies studied in this paper, only two are known to be Seyfert galaxies: NGC 3627 and NGC 1672 \citep{AGN_catalogue}. No correlation is seen between the presence of an AGN within these local galaxies and a larger or flatter semivariogram. Furthermore, the fraction of gas that would need to be lost to AGN winds in order increase the variance in the \citetalias{KT18} model by a factor of $\gtrsim 600$ is $\geq 96\%$, which is significantly larger than the amount of gas ejected by galactic winds observed in simulations of similar galaxies \citep[e.g.][]{Ma+16, Christensen+18}. Therefore, it is unlikely that AGN feedback is the mechanism responsible for the discrepancy in normalisation of the semivariograms seen between the data and the model. 

The \citetalias{KT18} model has the advantage of being simple enough to facilitate the computation of a theoretical semivariogram analytically. This simplicity is only possible because the model makes several simplifying assumptions, one or more of which may act to decrease the amplitude of predicted metallicity fluctuations. The assumption that all galaxies have a constant star formation history may lead to an underestimation of the degree of metal inhomogeneity within a starburst galaxy, as more recent metal injection events will have had less time to diffuse throughout the ISM, making them more strongly correlated.
When modelling the amount of metals added per star forming event, $\sigma_w^2$ is assumed to depend solely on variations in cluster mass -- yet modelling efforts by various teams, such as \citet{Nomoto+13}, show that supernova yields depend both on the mass of the star and on the metallicity of the progenitor star, in a highly nonlinear way. Similarly, the assumption that global galaxy parameters such as $h$ and $\sigma_g$ are constant throughout a galaxy may impact the inferred homogeneity of the ISM. To solve this issue, more covariates could be added to the model of the spatially-varying mean, $\mu(\vec{x})$. Possible covariates that can be added include any physical properties that have been observed to be correlated with metallicity on small scales, such as the stellar mass density \citep{Erroz-Ferrer+19} or star formation rate of each H{\sc ii} region \citep{Sanchez-Menguiano+19}, or an indicator function that differentiates between stellar arms and inter-arm regions \citep{Sanchez-Menguiano+20}. This data will be made available for the full set of PHANGS galaxies in an upcoming data release (Kreckel, K., private communication), facilitating such an analysis. We emphasise that the flexibility of the approach we have outlined in this paper to accommodate any models of the process mean and small-scale variance in a way that allows consistent comparisons to be drawn between different models is a very attractive feature of the geostatistical framework.

The predicted metallicity correlation structure of the \citetalias{KT18} model has been compared to IFS data in previous works \citep{Kreckel+20, Li+21}, albeit using different methodologies. In this work and in \citet{Kreckel+20}, the metallicity gradient is fit using a least-squared method and subtracted, whereas in \citet{Li+21} radial trends are removed by subtracting the median metallicity in $0.2$ kpc wide annuli. Their approach has the advantage of better capturing non-linear metallicity trends, at the cost of introducing discontinuities in the residual metallicity field at the boundaries between annuli. Further, using too many bins may result in an overfitting to the radial variations, washing out any true fluctuations in the radial direction. This would be incompatible with our geostatistical approach, as we need to assume that the field of residual metallicity fluctuations $\eta(\vec{x})$ is isotropic in order to predict the structure of the semivariogram from the two-point correlation function.

While \citet{Li+21} fit a two-point correlation function to the residual metallicity fluctuation map, in this work, a semivariogram is computed. This approach affords two advantages over the method of \citet{Li+21}: firstly, with the semivariogram, it is easier to correct for the variance caused by measurement error (in \citealt{Li+21}, the effects of measurement error complicate the correlation function, adding two new parameters to the \citetalias{KT18} model). Secondly, the semivariogram is sensitive to the magnitude of metallicity fluctuations, whereas the two-point correlation function is not. This additional degree of freedom revealed a tension between the model and the data that the two-point correlation functions could not identify.
Finally, in this work, the values of the parameters of the \citetalias{KT18} model are computed a priori from global galaxy properties, whereas in \citet{Li+21}, the parameters are fit to best reproduce the observed two-point correlation function for each galaxy. Because of this, some computed parameters of best fit are inconsistent with the physical properties of local galaxies, such as star-formation timescales of $\lesssim 1$ Gyr. 

In \citet{Kreckel+20}, after subtracting the overall metallicity gradients, the mean standard deviation of the scatter around the metallicity gradient is assessed for spaxels separated by less than some critical value. Their Figure 4 plots how this statistic increases as this separation is raised. These plots are intended to capture the same feature that is revealed by the semivariogram, and in this respect they are successful and yield consistent results. Yet, we highlight the advantages of the semivariogram approach introduced here for astronomical applications, as it is widely supported by a vast geostatistical literature, and has several useful mathematical properties, including the ability to separate correlated from uncorrelated sources of error (see Figure \ref{fig:svg}). \citet{Kreckel+20} also compute the two-point correlation function for the fluctuations of the metallicities about the linear gradient, reporting that ``most galaxies appear to have reached their global scatter by $\sim 3$ kpc scales", again in good agreement with our results. The $30\%$ and $50\%$ correlation scales are also reported, however the motivation for focusing on these percentiles is not discussed. The statistical framework presented in this work goes beyond that analysis, providing information not only about the spatial scale of metallicity fluctuations, but also about their amplitude in a rigorous way, underpinned by a robust statistical framework for the analysis of the data. Furthermore, the semivariogram method can be used to isolate these small-scale variations from the random uncorrelated fluctuations associated with uncertainties originating from measurement error or imprecise diagnostics.

In the \citetalias{KT18} model, turbulence is treated as a source of diffusion, following a mathematical model given in \citet{Karlsson+13}. \citet{Pan+Scan10} argue that such approximations overestimate the efficiency of turbulence as a source of mixing. Using a numerical solver to simulate a 1kpc$^2$ patch of ISM, \citet{deAvillez+MacLow02} show that accounting for the effects of turbulence increases mixing timescales, allowing small scale inhomogeneities to persist for long periods of time. Accounting for this effect would increase the amount of inhomogeneities found with respect to the \citetalias{KT18} model, as more recent metal enrichment events would not have had the time to break up into small regions, preventing them from diffusing into the ISM. A more realistic treatment of turbulence may explain not only the normalisation discrepancy in the semivariograms between the \citetalias{KT18} model and observations, but also the increased power of small-scale correlations seen in Figure \ref{fig:theory_svgs} for NGC 628 and NGC 2835.

In future works, we intend to extend our analysis to a suite of hydrodynamic simulations, including EAGLE \citep{EAGLE} and IllustrisTNG \citep{TNG1, TNG2, TNG3, TNG4}, in order to understand how differences in the sub-grid physical models implemented in these simulations lead to different small-scale metallicity structures \citep{Rennehan21}. Such an analysis could provide a new avenue along which simulation results could be compared to real-world data, resulting in improved constraints on processes such as supernova feedback and turbulent transport.

\section{Summary and conclusions} \label{sec:conclusions}

In this paper, we have shown that the techniques of geostatistics can be valuable for analysing high-resolution galaxy observation data; and can be used to inform and constrain models of galaxy evolution. We summarise our main results below:

\begin{itemize}
    \item The high-resolution IFU data captured by PHANGS-MUSE facilitates a multi-scale analysis of the metallicity structure of the ISM, and allows testing of predictions of metal-transport models, such as the analytical model of \citetalias{KT18}.
    \item For $7$ local star-forming main sequence galaxies with $9.6 \leq \log( M_*/M_\odot) \leq 10.6$, a semivariogram was computed, revealing the size and scale of metallicity fluctuations around the mean metallicity gradient.
    \item The metallicities of H\textsc{ii} regions separated by more than $\sim 1-3$ kpc were found to be uncorrelated, broadly consistent with the predictions of \citetalias{KT18}. No strong correlations were observed between the scale over which metallicities were correlated, and other global galaxy properties.
    \item From the difference between the values of the semivariograms at small and large separations, we observe approximately $50
    \%$ of the variance in metallicity throughout a galaxy to be correlated at small scales. Such correlation cannot be attributed to measurement error, and therefore must be a true feature of these star-forming galaxies. The amount of correlated variance seen within metallicity maps in the PHANGS galaxy sample is several orders of magnitude higher than the predictions of \citetalias{KT18} (see Section \ref{ssec:v_theory}).
    \item This discrepancy cannot be ascribed to measurement error in the global properties or local metallicity measurements of the galaxy sample. The most likely explanation is that the simplified model of \citetalias{KT18} leaves out some important features of metal transport (see Section \ref{sec:discussion}). In particular, the treatment of turbulence in the model of \citetalias{KT18} may lead to an underprediction of the size of small-scale metallicity fluctuations.
\end{itemize}
In the current work, we have focused on using geostatistics to gain an understanding of the small-scale metallicity structure of galaxies. Another application of geostatistical methods is to predict values of a random field at unmeasured points \citep[e.g.][]{Matheron69, UK}. This could be particularly valuable for studying metallicity variations throughout regions of diffuse ionised gas (DIG), where strong emission line diagnostics have not been calibrated and may lead to biased or inaccurate results \citep{Kreckel+19, Kumari+19}. We will demonstrate how metallicity maps of H\textsc{ii} regions may be interpolated using optimal, unbiased geostatistical algorithms to produce metallicity predictions for DIG-dominated regions in a future work.

While this series of papers focuses on metallicity,
the methods and models of spatial statistics outlined in this paper could also be used to understand multi-scale spatial variations in other galaxy properties, given high-resolution IFS data. We encourage other scientists to consider using these methods to study internal changes in other galaxy properties, such as the star formation rate, gas density, or temperature, in order to better understand the state and structure of the ISM.


\section*{acknowledgements}
We are grateful to the anonymous referee for their constructive and informative feedback, which contributed to improve the quality and readability of this paper.
The authors would further like to thank Prof. Mark Krumholz for several insightful clarifying discussions on the theoretical model investigated in this work; Ismael Pessa for providing determinations of the average SF densities within the galaxy sample; Dr. Leonid Pilyugin for providing data on the uncertainties associated with the Scal diagnostic; and Dr. Kathryn Kreckel for her insight regarding the metallicity gradients computed for the PHANGS sample.
BM acknowledges support from an Australian Government Research Training Program (RTP) Scholarship. This research is supported in part by the Australian Research Council Centre of Excellence for All Sky Astrophysics in 3 Dimensions (ASTRO 3D), through project number CE170100013.

\section*{data availability}
All PHANGS data used in this analysis is publically available for download at \url{www.phangs.org/data}. Any further data products created for this work are available from the corresponding author upon reasonable request.

\bibliographystyle{mnras}
\bibliography{biblio} 

\newpage
\appendix

\section{Analysis with an alternative metallicity diagnostic}
\label{ap:O3N2}

Different metallicity diagnostics often yield partially inconsistent results: not only in terms of their normalisation \citep{Kewley+Ellison08}; but also in their range \citep[e.g.][]{Erroz-Ferrer+19}, and on small-scales by some nonlinear factor \citep{Kreckel+19, Li+21}. Given that here we are only investigating deviations from the mean metallicity within galaxies, any change in normalisation from one galaxy to another will not affect our results. However, changes in the range of metallicities inferred through the use of each diagnostic may affect our results, as well as any nonlinear diagnostic errors. Thus, in this section, we repeat our analysis using an alternative strong line metallicity diagnostic: the O3N2 diagnostic, together with the calibration of \citet{Marino+13}. We choose this calibration because it is simple to implement, widely used, and can be computed from the emission line data made publically available in \citet{Kreckel+19}.

\begin{figure*}
    \centering
    \includegraphics[width=0.9\textwidth]{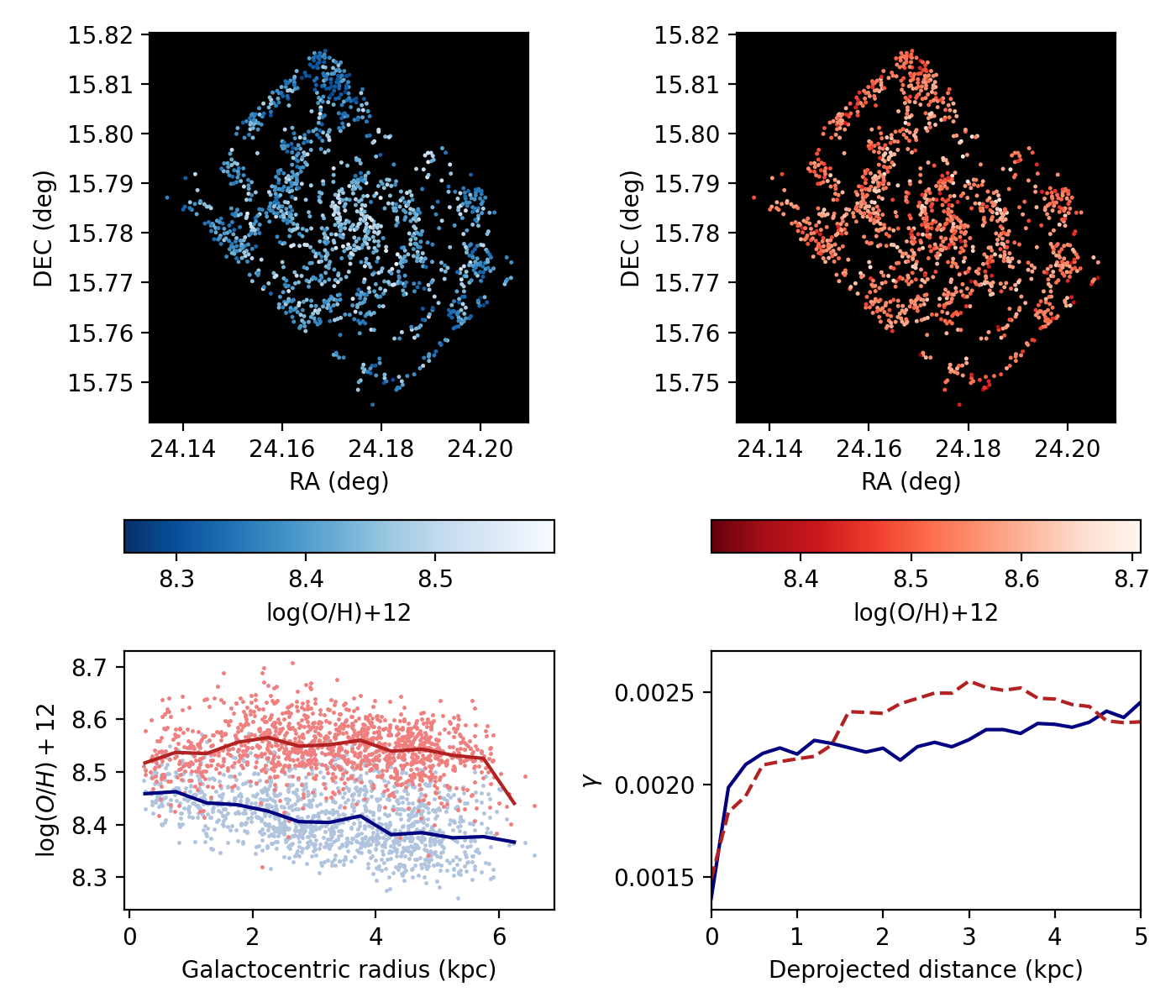}
    \caption{ \emph{Top left:} A map of H\textsc{ii} region metallicities computed for NGC 682 using the Scal diagnostic, with the calibration of \citet{Pilyugin+Grebel16}. \emph{Top right:} The same H\textsc{ii} region map, but with metallicities computed using the O3N2 diagnostic with the calibration of \citet{Marino+13}. \emph{Bottom left:} Comparing the median radial metallicity profiles for these diagnostics. Metallicities obtained using the O3N2 diagnostic (red) tend to be higher than those obtained using Scal (blue). In this galaxy, in the central $\sim 2$ kpc, the metallicity gradient is seen to be inverted when using the O3N2 diagnostic, but not the Scal diagnostic. \emph{Bottom right:} Empirical semivariograms of the residual error around a linear metallicity gradient model for the Scal diagnostic (blue solid line) and the O3N2 diagnostic (red dashed line). The rise in the O3N2 semivariogram between $2$ and $4$ kpc can be attributed to the failure of the simple metallicity gradient model to account for the inverted metallicit gradient seen in the inner $2$kpc.
    }
    \label{fig:O3N2_v_Scal_NGC628}
\end{figure*}

\begin{figure*}
    \centering
    \includegraphics[width=0.9\textwidth]{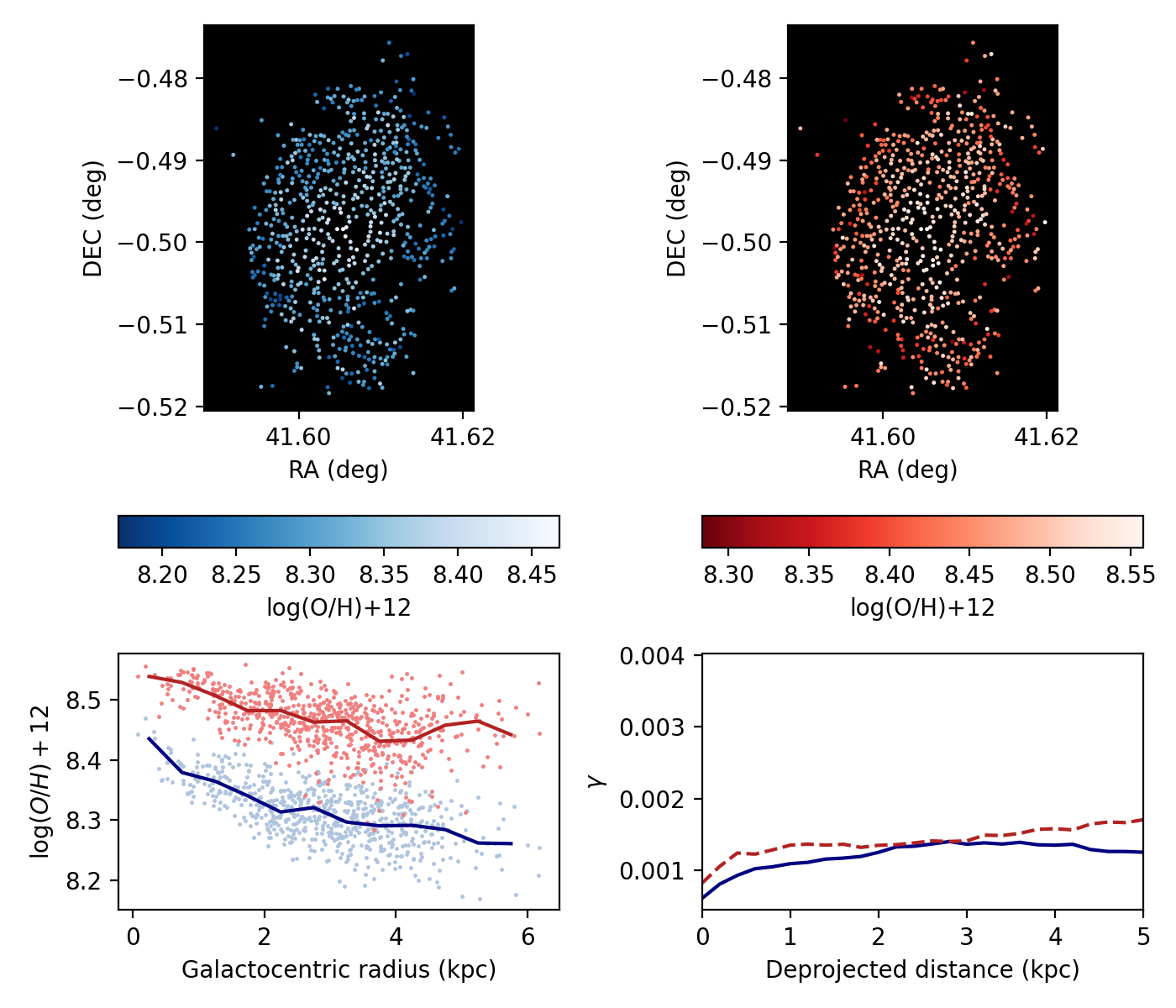}
    \caption{As \ref{fig:O3N2_v_Scal_NGC628}, but for NGC 1087.
    }
    \label{fig:O3N2_v_Scal_NGC1087}
\end{figure*}

\begin{figure*}
    \centering
    \includegraphics[width=0.9\textwidth]{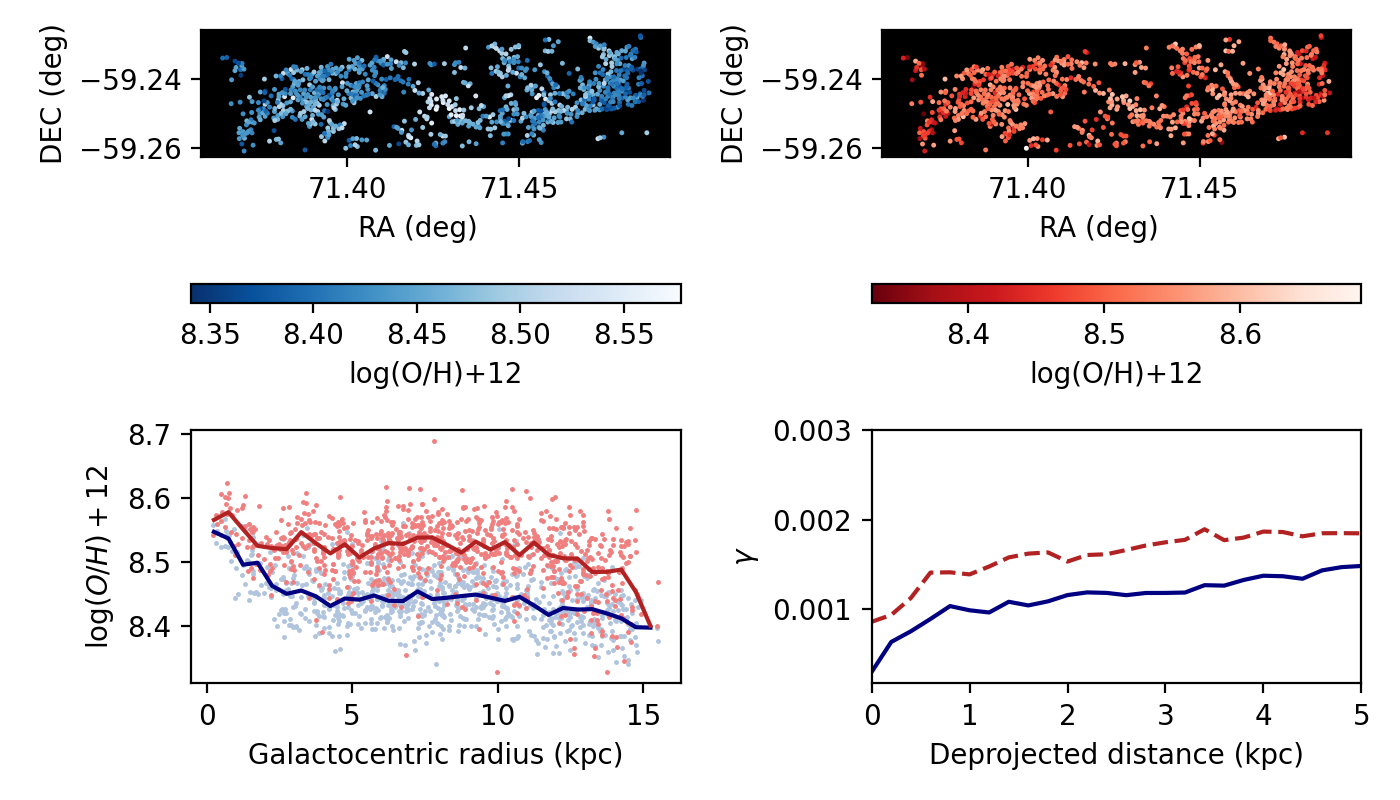}
    \caption{As \ref{fig:O3N2_v_Scal_NGC628}, but for NGC 1672.
    }
    \label{fig:O3N2_v_Scal_NGC1672}
\end{figure*}

\begin{figure*}
    \centering
    \includegraphics[width=0.9\textwidth]{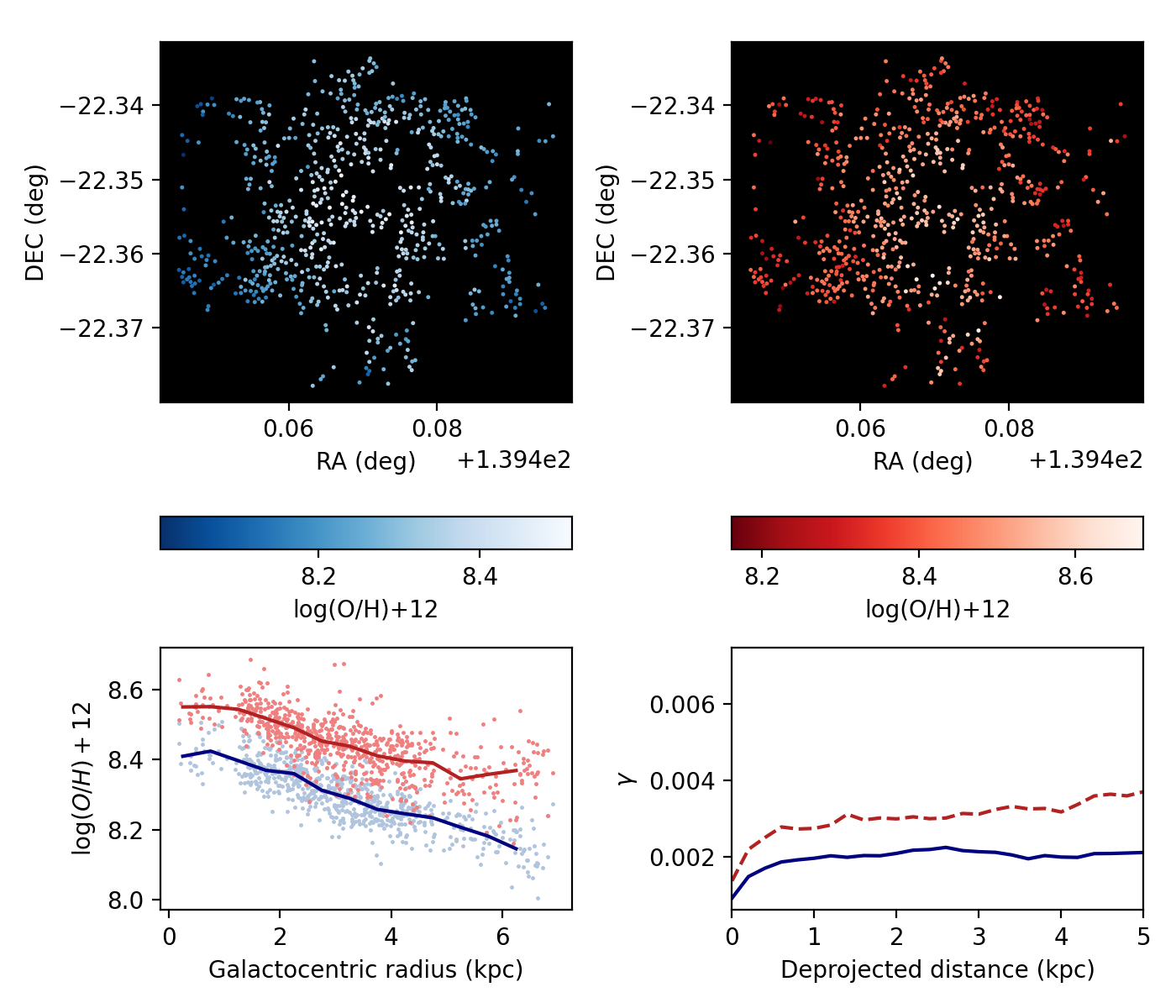}
    \caption{As \ref{fig:O3N2_v_Scal_NGC628}, but for NGC 2835.
    }
    \label{fig:O3N2_v_Scal_NGC2835}
\end{figure*}

\begin{figure*}
    \centering
    \includegraphics[width=0.9\textwidth]{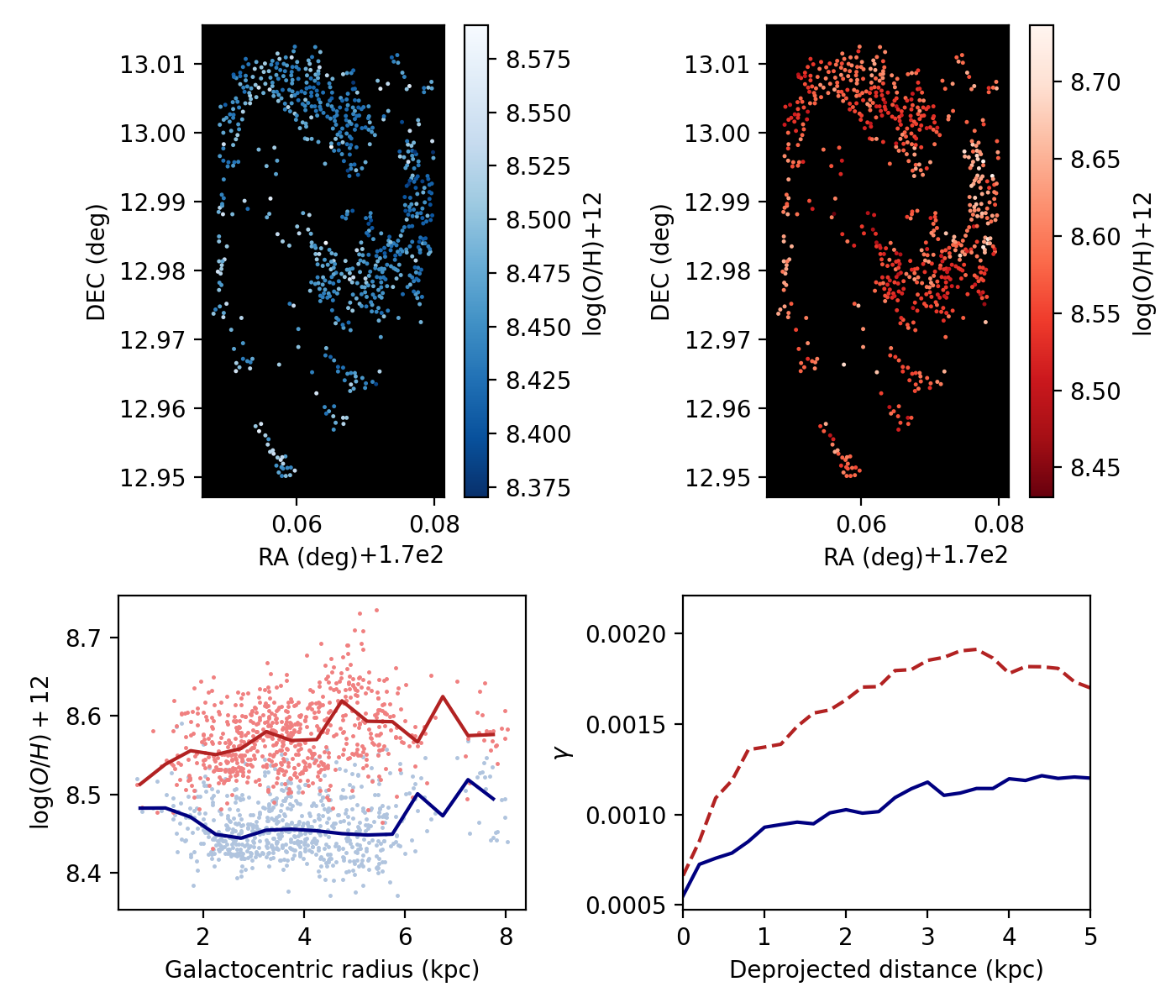}
    \caption{As \ref{fig:O3N2_v_Scal_NGC628}, but for NGC 3627.
    }
    \label{fig:O3N2_v_Scal_NGC3627}
\end{figure*}

\begin{figure*}
    \centering
    \includegraphics[width=0.9\textwidth]{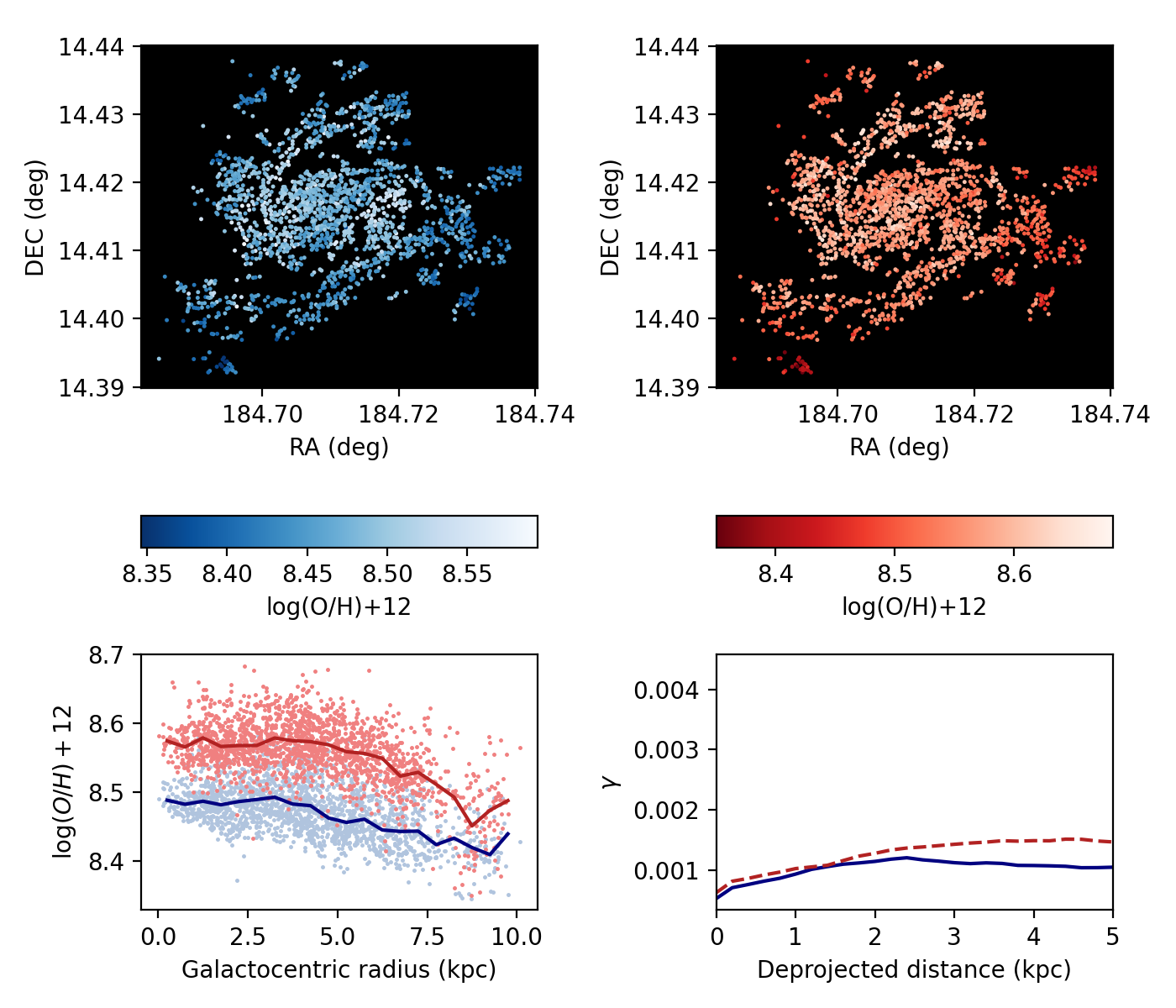}
    \caption{As \ref{fig:O3N2_v_Scal_NGC628}, but for NGC 4254.
    }
    \label{fig:O3N2_v_Scal_NGC4254}
\end{figure*}

\begin{figure*}
    \centering
    \includegraphics[width=0.9\textwidth]{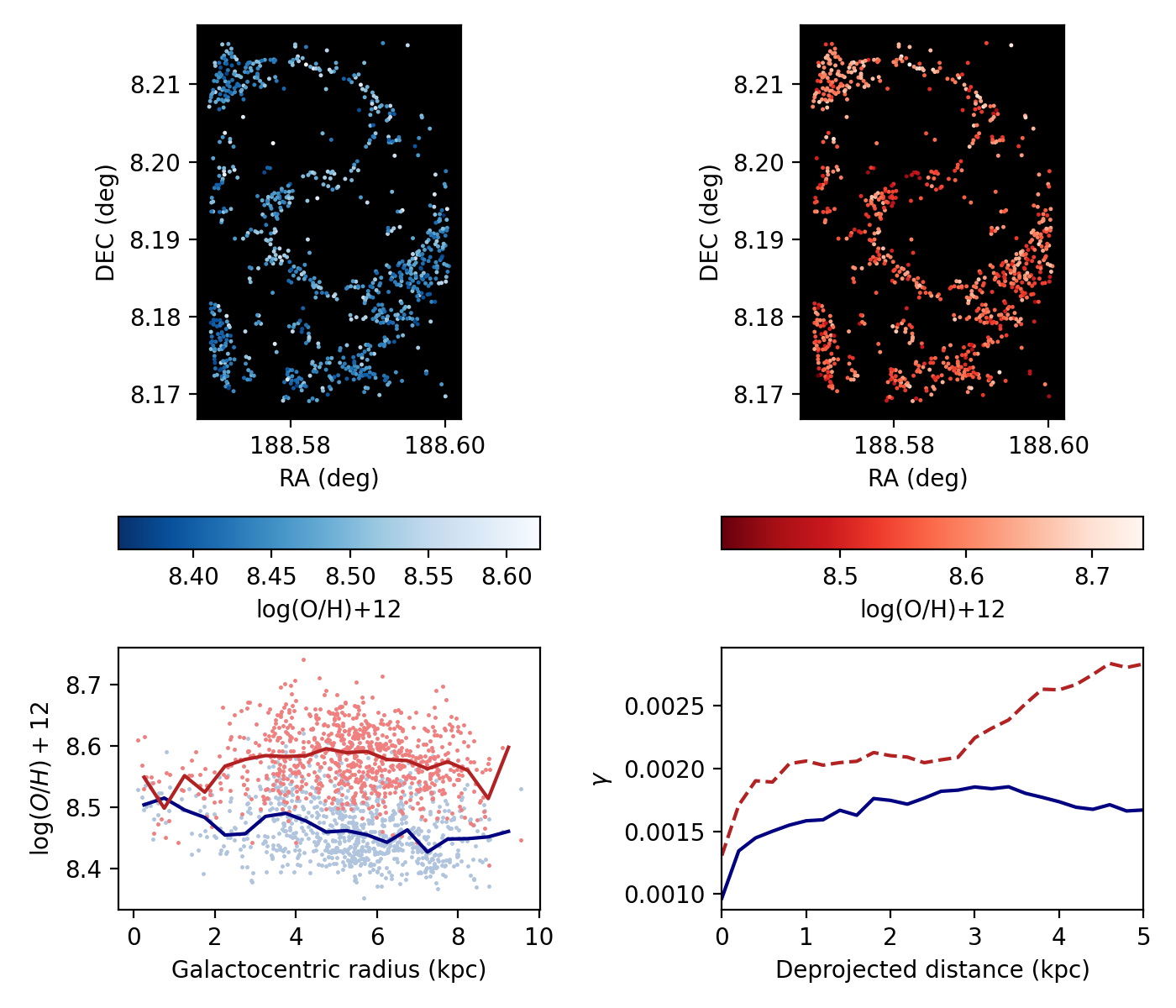}
    \caption{As \ref{fig:O3N2_v_Scal_NGC628}, but for NGC 4535. Here, the rise in the empirical semivariogram calculated via the O3N2 diagnostic (bottom right panel, dashed red line) between 3kpc and 6kpc can be attributed to the inverted metallicity gradient seen when using this diagnostic in the inner 3 kpc (bottom left panel, solid red line).
    }
    \label{fig:O3N2_v_Scal_NGC4535}
\end{figure*}

Empirical semivariograms for these galaxies computed with this diagnostic are compared to those computed using the Scal diagnostic together with the calibration of \citet{Pilyugin+Grebel16} in Figures \ref{fig:O3N2_v_Scal_NGC628}-\ref{fig:O3N2_v_Scal_NGC4535}. For NGC 1672, NGC 2835, and to some extent NGC 3627, the shape of the empirical semivariograms computed using the O3N2 diagnostic matches the shape of the semivariogram computed using the Scal diagnostic, with the overall values of the semivariogram translated upwards by a constant factor (typically $4\times 10^{-4}-8\times 10^{-4}$). This reflects the additional measurement uncertainty associated with the O3N2 diagnostic: for NGC 1672, this corresponds to an additional uncertainty of $\sigma_Z = 0.03$ dex, whereas for NGC 2835, this corresponds to an additional uncertainty of $\sigma_Z = 0.04$ dex. The size of the offset between the O3N2 semivariogram and the Scal semivariogram for each galaxy is comparable to the height of
the Scal semivariogram at null separation, indicating that the error associated with the O3N2 diagnostic is about twice that of the Scal diagnostic. This is consistent with comparisons of both diagnostics to auroral line based measurements of metallicity: in \citet{Marino+13}, the scatter of the relationship between the metallicity computed using the O3N2 method and the direct method was shown to be $\sigma_{O3N2} = 0.18$ dex, whereas for the S-calibration of \citet{Pilyugin+Grebel16}, the scatter between the metallicity inferred from this strong-line diagnostic and the $T_e$-based metallicity measurement is $\sigma_{Scal} = 0.048$ dex.

Interestingly, \citet{Kreckel+19} showed that small-scale metallicity fluctuations about the mean metallicity gradient measured using these two diagnostics showed no correlation (see Figure 22 of Appendix C). Recast in the language of our model, they showed that $\epsilon_{\rm Scal}(\vec{x})+\eta_{\rm Scal}(\vec{x})$ was largely independent of $\epsilon_{\rm O3N2}(\vec{x})+\eta_{\rm O3N2}(\vec{x})$. From the similar shape present in their semivariograms, we see that the small scale fluctuations recovered using either diagnostic are very similar for most galaxies -- that is, $\eta_{\rm O3N2}(\vec{x}) \approx \eta_{\rm Scal}(\vec{x})$. Therefore, the lack of correlation seen in \citet{Kreckel+19} must be caused by an independence in the errors associated with each diagnostic -- that is, $\epsilon_{\rm Scal}(\vec{x})$ and $\epsilon_{\rm O3N2}(\vec{x})$ are uncorrelated at each location $\vec{x}$, and the size of these uncertainties is large in comparison to the small-scale metallicity fluctuations $\eta(\vec{x})$ that we are able to isolate and detect using semivariograms.

For NGC 628 and NGC 4535, large fluctuations are seen in the empirical semivariogram computed using the O3N2 diagnostic. These fluctuations come from a failure of a simple linear metallicity gradient model to capture the large-scale behaviours of these galaxies. For NGC 628, using the O3N2 diagnostic, at distances larger than $2$ kpc, the metallicity gradients inferred from the O3N2 diagnostic agree with those inferred from the Scal diagnostic. However, at smaller radii, the metallicity gradient is inverted. This has been seen before for NGC 628 \citep{Rosales-Ortega11, Sanchez+11} and NGC 4535 \citep{Sanchez-Menguiano+18} when the O3N2 diagnostic is used, and is not uncommon for disk galaxies with star-forming rings \citep[e.g.][]{Sanchez+14}. The unusual structure of this galaxy with a low-metallicity centre appears in the semivariogram as an increased amount of variance on the scales of $\sim 3 $kpc, roughly the spatial scale of this low-metallicity central region. A similar behaviour is seen in the metallicity profile of NGC 4535: in the central $\sim 3$ kpc, metallicity gradients are inverted, leading to a broad rise in the semivariogram at a separation of $\sim 6$ kpc. Interestingly, such behaviours are not seen when the Scal diagnostic is used, but investigating the origin of this is beyond the scope of our analysis..

Despite these considerations, in all cases, the amount of correlated variance in the metallicity measured by the O3N2 diagnostic is always greater than or equal to the amount of correlated variance measured using the Scal diagnostic. This means that the choice of diagnostic does not affect our main conclusion that the metal transport model of \citetalias{KT18} underpredicts the amount of correlated variance in metallicity.

\label{lastpage}
\end{document}